%% file: main.tex
\documentclass[twocolumn,trackchanges]{aastex631}
\input{preamb.tex}

\shorttitle{Generative Models of Multi-channel Dust Emission Maps}
\shortauthors{Régaldo-Saint Blancard et al.}

\begin{document}

%ORCID ?

\author[0000-0003-0055-0953]{Bruno Régaldo-Saint Blancard}
\correspondingauthor{Bruno Régaldo-Saint Blancard}
\email{bregaldosaintblancard@flatironinstitute.org}
\affiliation{Center for Computational Mathematics, Flatiron Institute, 162 5\textsuperscript{th} Avenue, New York, NY 10010, USA}

\author{Erwan Allys}
\affiliation{Laboratoire de Physique de l’ENS, ENS, Université PSL, CNRS, Sorbonne Université, Université Paris Cité, 75005 Paris, France}

\author{Constant Auclair}
\affiliation{Laboratoire de Physique de l’ENS, ENS, Université PSL, CNRS, Sorbonne Université, Université Paris Cité, 75005 Paris, France}

\author[0000-0003-1097-6042]{François Boulanger}
\affiliation{Laboratoire de Physique de l’ENS, ENS, Université PSL, CNRS, Sorbonne Université, Université Paris Cité, 75005 Paris, France}

\author{Michael Eickenberg}
\affiliation{Center for Computational Mathematics, Flatiron Institute, 162 5\textsuperscript{th} Avenue, New York, NY 10010, USA}

\author[0000-0002-3065-9944]{François Levrier}
\affiliation{Laboratoire de Physique de l’ENS, ENS, Université PSL, CNRS, Sorbonne Université, Université Paris Cité, 75005 Paris, France}

\author[0000-0001-9551-1417]{Léo Vacher}
\affiliation{IRAP, Université de Toulouse, CNRS, CNES, UPS, Toulouse, France}

\author{Sixin Zhang}
\affiliation{Université de Toulouse, INP, IRIT, Toulouse, France}

\title{Generative Models of Multi-channel Data from a Single Example - Application to Dust Emission}

\input{abstract.tex}


% Body
\input{intro.tex}
\input{data.tex}
\input{methodo.tex}
\input{models.tex}
\input{conclusion.tex}

% Acknowledgements
\input{acknowledgements.tex}

% Bibliography
\bibliographystyle{aasjournal}
\bibliography{bib, planck_bib}

% Appendices
\input{appendices.tex}

\end{document}

%% file: preamb.tex
\usepackage{amsmath, mathtools}
\usepackage{amssymb} % for \lessapprox notably
\usepackage{amsfonts,bm,bbm}
\usepackage{stmaryrd}
\usepackage{graphicx}
\usepackage{upgreek}
\usepackage{comment}

\renewcommand*\vec[1]{\ensuremath{\boldsymbol{#1}}}
\newcommand{\ve}[1]{{\vec{#1}}}
\newcommand{\norm}[1]{\left\lVert#1\right\rVert}

\newcommand\dI{\mathrm{d}}

\newcommand*\E[1]{\mathrm{E}\left( #1 \right)}

\newcommand*\Cov[2]{\mathrm{Cov}\left( #1, #2 \right)}
\newcommand*\Covb[2]{\mathrm{Cov}\left[ #1, #2 \right]}

\newcommand*{\conj}[1]{#1^*}
\newcommand*{\conjb}[1]{\overline{#1}}

\providecommand{\sorthelp}[1]{} % For Planck_bib.bib

%% file: abstract.tex
\begin{abstract}

The quest for primordial $B$-modes in the cosmic microwave background has emphasized the need for refined models of the Galactic dust foreground. Here, we aim at building a realistic statistical model of the multi-frequency dust emission from a single example. We introduce a generic methodology relying on microcanonical gradient descent models conditioned by an extended family of wavelet phase harmonic (WPH) statistics. To tackle the multi-channel aspect of the data, we define \textit{cross-WPH} statistics, quantifying non-Gaussian correlations between maps. Our data-driven methodology could apply to various contexts, and we have updated the software \texttt{PyWPH}, on which this work relies, accordingly. Applying this to dust emission maps built from a magnetohydrodynamics simulation, we construct and assess two generative models of: 1) a $(I,E,B)$ multi-observable input, 2) a $\{I_\nu\}_\nu$ multi-frequency input. The samples exhibit consistent features compared to the original maps. A statistical analysis of 1) shows that the power spectra, distributions of pixels, and Minkowski functionals are captured to a good extent. We analyze 2) by fitting the spectral energy distribution (SED) of both the synthetic and original maps with a modified blackbody (MBB) law. The maps are equally well fitted, and a comparison of the MBB parameters shows that our model succeeds in capturing the spatial variations of the SED from the data. Besides the perspectives of this work for dust emission modeling, the introduction of cross-WPH statistics opens a new avenue to characterize non-Gaussian interactions across different maps, which we believe will be fruitful for astrophysics.

\end{abstract}

%% file: intro.tex
\section{Introduction}

Since the \textit{Planck} mission~\citep{planck2016-l01}, the observation of the cosmic microwave background (CMB) has become closely entwined with the physics of the interstellar medium (ISM). On the one hand, the thermal emission of our own Galaxy severely contaminates the CMB signal, acting as a foreground to the primordial signal. On the other hand, CMB experiments provide unique data for interstellar astrophysics. Component separation methods~\citep[see e.g.][]{planck2016-l04}, which aim at recovering the accurate CMB maps, have to deal with the non-Gaussian\footnote{We call ``non-Gaussian", any statistical feature that is not characterized by the power spectrum (or cross spectrum, when considering a pair of fields).} structure of the dust emission and the spatial variations of the spectral energy distribution (SED). In the quest for primordial $B$-modes~\citep{Kamionkowski2016} in the CMB polarization signal, which is one of the main targets of the new generation of CMB experiments such as ACTPol~\citep{Naess2014}, SPIDER~\citep{Fraisse2013}, LiteBIRD~\citep{Ishino2016}, the Simons Observatory~\citep{Simons-Observatory-2019}, and CMB-S4~\citep{CMB-S4-Science-Case}, high-precision models of the multi-frequency polarized dust foreground have become crucial~\citep{pb2015}.

In this paper, we address the problem of building a statistical model of multi-frequency dust emission maps by introducing a generic methodology that does not rely on any prior phenomenological model (i.e. a purely \textit{data-driven} approach). We aim at taking into account the highly non-Gaussian properties of the data, and modeling the correlations between the frequency channels and between the polarization and total intensity observables. Because, from an observational point of view, we only have a single sky, we also choose to build our model from a single multi-channel input. In other words, if we call this input $x$, the goal is to approximate the distribution of the underlying random field $X$ based on the single realization $x$. Approximating a high-dimensional distribution is always a daunting task, and with this additional constraint this becomes even harder if not intractable. To partially alleviate this difficulty, we will thus assume (spatial) statistical homogeneity for $X$. Finally, we want our model to be \textit{generative}, in the sense that drawing new samples of the model should be doable in a reasonable amount of time. This might be a crucial constraint for component separation methods requiring an important amount of simulated maps of dust emission, such as methods adopting a simulation-based inference approach as in~\cite{Jeffrey2021}.

Various models of the polarized emission of dust have been developed in the literature~\citep[for an extensive discussion, see][Chapter~II]{RegaldoThese2021}, some of them being packed in CMB sky simulator softwares~\citep[e.g.][]{Delabrouille2013, Thorne2017}. Among these models, we make a distinction between deterministic approaches, designed to retrieve the true emission of the sky~\citep[e.g.][]{planck2013-p06b, planck2014-a12, planck2016-XLVIII, planck2016-l04}, and statistical approaches, focusing on its statistical properties~\citep[e.g.][]{Vansyngel2017, Hervias2022}. We also distinguish phenomenological approaches, making use of phenomenological priors (e.g. modified blackbody (MBB) SED), from data-driven approaches. Our approach stands in the class of statistical and data-driven models. In this category, \cite{Aylor2021}, and then \cite{Thorne2021}, employed a generative adversarial network (GAN) and a variational autoencoder (VAE), respectively, to model the total intensity of mono-frequency dust emission maps. \cite{Krachmalnicoff2021} made use of a GAN to generate sub-degree angular scales in mono-frequency dust polarization maps. These approaches, which involve techniques from the emerging field of deep generative modeling, rely on convolutional neural networks (CNNs) that need to be trained. Such training steps require at least hundreds if not thousands of observations, which obviously bind the models to their respective training sets. On the other hand, \cite{Allys2019} and then \cite{Regaldo2020} introduced generative models of dust emission maps in total intensity and polarization, respectively, that do not necessitate any training and can be built from a single input map. These models are approximate maximum entropy models, called \textit{microcanonical gradient descent models}, that are conditioned by wavelet scattering transform (WST) moments~\citep{Mallat2012, Bruna2013, Bruna2019}. \cite{Jeffrey2021} made use of a similar kind of model for dust polarization data but using wavelet phase harmonic (WPH) moments~\citep{Mallat2020, Zhang2021, Allys2020}. Note that although the WST and WPH statistics share many similarities with the representations learned by CNNs~\citep{Mallat2016, Mallat2020}, they avoid some of their drawbacks, namely the lack of interpretability and the necessity of training. However, until now, none of these existing generative models have tackled the multi-channel aspect of dust emission data (besides the joint modeling of the linear polarization variables, see e.g.~\cite{Regaldo2020}). We propose to bridge this gap by incorporating a modeling of correlations between maps at different frequency channels as well as correlations between total intensity and polarization maps.

Observations of dust polarized emission at a given frequency channel $\nu$ take the form of a triplet of two-dimensional Stokes parameter maps $(I_\nu, Q_\nu, U_\nu)$, with $I_\nu$ the total intensity of the emission and $(Q_\nu, U_\nu)$ describing the linear polarization signal. For CMB science, the polarization maps $(Q_\nu, U_\nu)$ are usually transformed into $(E_\nu, B_\nu)$ maps~\citep{Zaldarriaga2001}. In order to make the present work more readily usable in this cosmological context, we adopt the same polarization variables in the following.

In the continuity of the above series of works, we introduce microcanonical gradient descent models of multi-channel dust emission maps $\{(I_\nu, E_\nu, B_\nu)\}_{\nu}$ conditioned by a new family of WPH statistics. In particular, we introduce \textit{cross}-WPH statistics, allowing us to combine multiple maps and characterize their non-Gaussian correlations. This paper focuses on the methodology and the validation of such models. We build two models demonstrating the capabilities of our approach: the first one is a mono-frequency model based on a triplet of simulated dust maps $(I_{\nu_0}, E_{\nu_0}, B_{\nu_0})$, the second one is a multi-frequency model of a set of total-intensity simulated maps $\{I_\nu\}_{\nu}$. Finally, let us emphasize that, although the focus here is on dust emission maps, our approach does not involve any prior model that is specific to dust data. As a consequence, our methodology could be applied in a very similar way to other astrophysical contexts, and more generally to other scientific contexts. All the necessary material to do so is publicly available within the Python package~\texttt{PyWPH}\footnote{\url{https://github.com/bregaldo/pywph/}}~\citep{Regaldo2021}.

This paper is organized as follows. In Sect.~\ref{sec:presentation_data}, we introduce the simulated multi-channel data that will be the target of our models. Then, in Sect.~\ref{sec:methodology}, we explain the underlying formalism of our approach. We introduce microcanonical gradient descent models and the family of WPH statistics that we will employ. In Sect.~\ref{sec:models} we present and validate our models. Finally, Sect.~\ref{sec:conclusion} summarizes our conclusions. This paper also includes three appendices. Appendix~\ref{app:microcanonical} presents maximum entropy microcanonical models, which underlie the definition of microcanonical gradient descent models. Appendix~\ref{app:app_wph} gives additional details on our family of WPH statistics. Appendix~\ref{app:app_gaussian} properly defines the Gaussian model that we will use as a baseline in Sect.~\ref{sec:monofreq_model}.

The data and codes to reproduce the models are provided on \url{https://github.com/bregaldo/dust_genmodels}.

%% file: data.tex
\section{Presentation of the data}
\label{sec:presentation_data}

In this section, we build a set of simulated multi-frequency maps $\{(I_\nu, E_\nu, B_\nu)\}_{\nu}$ of dust emission by proceeding as follows:
\begin{enumerate}
    \item In Sect.~\ref{sec:mdh_sim}, we extract a statistically homogeneous 3D gas density field and magnetic field from a magnetohydrodynamics (MHD) simulation designed to reproduce typical conditions of the diffuse ISM.
    \item In Sect.~\ref{sec:radiative_transfer}, we build Stokes maps $I_\nu$, $Q_\nu$, and $U_\nu$ based on the previous simulation and a phenomenological model of the SED of dust grains.
    \item In Sect.~\ref{sec:data_maps}, we transform the $\{(Q_\nu, U_\nu)\}_{\nu}$ maps into $\{(E_\nu, B_\nu)\}_{\nu}$ maps and show the resulting data in Fig.~\ref{fig:data_IEB}.
\end{enumerate}

\subsection{MHD simulation}
\label{sec:mdh_sim}

In order to consider a realistic gas density field $n$ and magnetic field $\ve{B}$, we make use of a MHD simulation designed to reproduce typical conditions of the diffuse ISM. This simulation is the same as the one used in \cite{Regaldo2020}. We briefly summarize its main characteristics in the following, and refer to this paper for additional details.

The simulation was run in the context of~\cite{Bellomi2020} to study the biphasic nature of the diffuse ISM. It employs the adaptive mesh refinement code {\tt RAMSES}~\citep{Teyssier2002,Fromang2006} to solve the equations of ideal MHD. The volume of the simulation is $(50~\mathrm{pc})^3$, and it is ultimately discretized on a $512^3$ mesh with periodic boundary conditions. Heating and cooling processes of the gas are taken into account, whereas self-gravity is neglected. An isotropic Habing radiation field with $G_0 = 1$ is applied at the boundaries of the box. An isotropic turbulent forcing is also applied, leading to a statistical steady state after a few turnover times. In this stationary regime, we have a velocity dispersion $\sigma_v \sim 2.6~\text{km}.\text{s}^{-1}$, the magnetic field has a mean component $\bar{\ve{B}} = B_0\ve{e}_z$ with $B_0 \sim 3.8~\upmu\text{G}$, and a dispersion $\sigma_B \sim 3.1~\upmu\text{G}$. Finally, the mean and dispersion of the gas density field $n$ are $\bar{n}=1.5~\text{cm}^{-3}$ and $\sigma_n \sim 4.7~\text{cm}^{-3}$, respectively.

We extract a snapshot in this stationary regime, which thus provides the gas density field $n$ and magnetic field $\ve{B}$ we were looking for.

\subsection{Stokes emission maps}
\label{sec:radiative_transfer}

For a given frequency $\nu$ and line of sight, and in the optically thin limit, the Stokes parameters $I_\nu$, $Q_\nu$, and $U_\nu$ can be expressed as follows~\citep[see e.g.][and references therein]{planck2014-XX}:
\begin{align}
	I_\nu &= \int S_\nu \left[1-p_0\left(\cos^2\gamma - \frac{2}{3}\right)\right]\mathrm{d}\tau_\nu,\label{eq:I_dust}\\
	Q_\nu &= \int p_0S_\nu \cos\left(2\phi\right)\cos^2\gamma\mathrm{d}\tau_\nu,\label{eq:Q_dust}\\
	U_\nu &= \int p_0S_\nu \sin\left(2\phi\right)\cos^2\gamma\mathrm{d}\tau_\nu,\label{eq:U_dust}
\end{align}
where $S_\nu$ is the source function of the dust emission, $\tau_\nu$ is the dust optical depth, $p_0$ is an intrinsic polarization fraction parameter, $\gamma$ is the angle that the local magnetic field makes with the plane of the sky, and $\phi$ is the angle that the projection of the local magnetic field on the plane of the sky makes with some arbitrary reference axis in this plane. Note that the particular choice of this reference axis does not impact the $(E_\nu, B_\nu)$ maps that will be derived in Sect.~\ref{sec:data_maps}. The infinitesimal dust optical depth reads ${\mathrm{d}\tau_\nu = \sigma_\nu n \dI l}$ where $\sigma_\nu$ is the dust cross section per H at frequency $\nu$, $n$ is the hydrogen density, and $\dI l$ is the infinitesimal element of length along the line of sight. All these quantities are a priori functions of the position on the line of sight.

We make the following simplifying assumptions. We assume that the polarization fraction parameter $p_0$, which is related to cross-section parameters of the grains and their degree of alignment with the magnetic field~\citep[for more details, see][]{planck2014-XX}, is uniform~\citep[for a discussion, see][]{Reissl2020}. Similarly to \cite{planck2014-XX}, we choose a typical value of $p_0 = 0.2$. We choose the source function of the grains $S_\nu$ to be that of a blackbody radiation $B_\nu(T)$, thus depending on the dust temperature $T$. We assume that the frequency dependence of the dust cross section follows a power law $\sigma_\nu = \sigma_0(\nu / \nu_0)^\beta$, where $\sigma_0$ is the dust cross section at the reference frequency $\nu_0$ and $\beta$ is the spectral index. This assumption is quite usual in the literature~\citep[see e.g.][and references therein]{planck2013-p06b}. We assume a uniform $\sigma_0$, whose precise value only affects the global normalization of the maps, and has no impact on the rest of this paper. Finally, we arbitrarily choose the $z$-axis of the simulation as the line of sight. Note that this axis corresponds to the direction of the mean magnetic field of the simulation, so that we expect statistical isotropy in the resulting 2D maps~\citep{Regaldo2020}.

The fields $T$ and $\beta$ are defined with respect to the local values of the density field $n$ using the following simple phenomenological model. The voxels of $T$ and $\beta$ are both exactly Gaussian distributed in a way that is consistent with the two following observational facts: 1) temperature is usually lower (higher) in higher (lower) density regions, 2) MBB fits of observational maps of the dust emission show that the fitted $T$ and $\beta$ parameters tend to be anti-correlated~\citep{planck2013-p06b}. Means and standard deviations of $T$ and $\beta$ are chosen to be [15~\text{K}, 1~\text{K}] and [1.5, 0.2], respectively. Formally, calling $\mathcal{R}(n(\ve{r}))$ the rank of a particular value $n(\ve{r})$ of the simulated density field, we thus define $T(\ve{r})$ and $\beta(\ve{r})$ as follows:
\begin{align}
    \beta(\ve{r}) &= F^{-1}_{\mathcal{N}(1.5, 0.2^2)}\left(\frac{\mathcal{R}(n(\ve{r}))}{1 + N^3}\right), \\
    T(\ve{r}) &= F^{-1}_{\mathcal{N}(15\mathrm{K}, \left(1\mathrm{K}\right)^2)}\left(1-\frac{\mathcal{R}(n(\ve{r}))}{1 + N^3}\right),
\end{align}
where $F^{-1}_{\mathcal{N}(m, \sigma^2)}$ is the inverse cumulative distribution function of a Gaussian distribution of mean $m$ and standard deviation $\sigma$, and $N = 512$ is the resolution of the simulation. Note that although this 3D phenomenological model of the emission of dust grains takes inspiration from MBB analyses of dust emission maps, there is a priori no reason for the SED of the projected maps $I_\nu$, $Q_\nu$, and $U_\nu$ to be amenable to a MBB~\citep[see e.g.][]{Chluba2017,McBride2022}.

\subsection{\texorpdfstring{$(E, B)$}{(E, B)} transform and resulting maps}
\label{sec:data_maps}

\begin{figure*}
    \centering
    \includegraphics[width=\hsize]{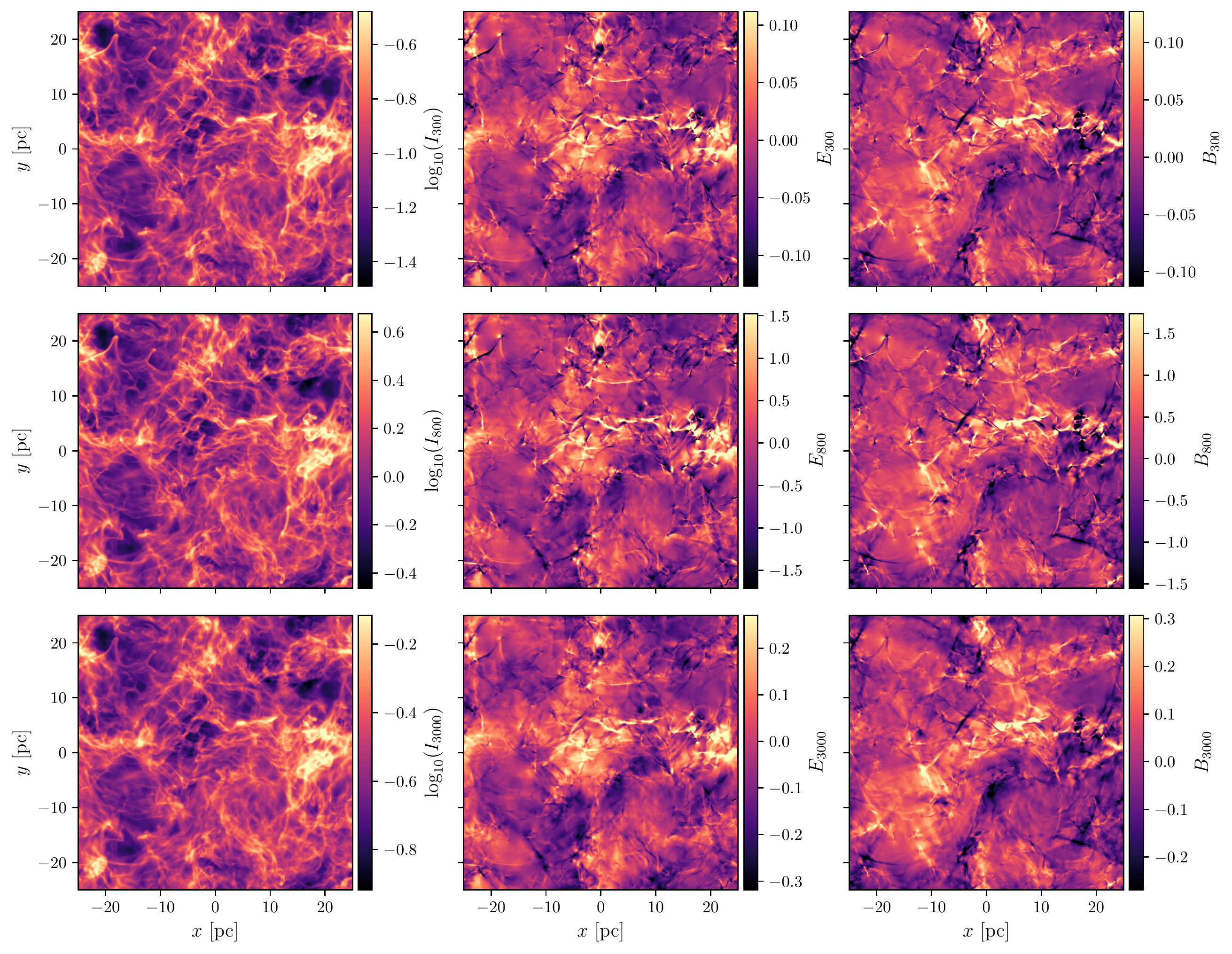}
    \caption{$\log_{10}(I_\nu)$, $E_\nu$, and $B_\nu$ (from left to right) simulated maps of dust emission at three different frequency channels: 300, 800, and 3000~GHz (from top to bottom).}
    \label{fig:data_IEB}
\end{figure*}

In a cosmological context, linear polarization is not often studied straight from the $Q$ and $U$ observables, but rather in terms of $E$ and $B$-modes~\citep{Zaldarriaga2001}. $E$ and $B$-modes are linear transforms of $Q$ and $U$ defined as follows in Fourier space and in the flat-sky approximation~\citep[see e.g.][]{Kamionkowski2016}:
\begin{align}
\begin{pmatrix} \hat{E}(\ve{k}) \\ \hat{B}(\ve{k}) \end{pmatrix} = \frac1{\sqrt{2}}\begin{pmatrix}\cos 2\alpha_\ve{k} & \sin 2\alpha_\ve{k}\\ -\sin 2\alpha_\ve{k} & \cos 2\alpha_\ve{k}\end{pmatrix} \begin{pmatrix}\hat{Q}(\ve{k}) \\ \hat{U}(\ve{k})\end{pmatrix},
\end{align}
where $\alpha_\ve{k}$ is the angle that $\ve{k}$ makes with the $x$-axis of the maps. The motivation for this transformation is twofold. First, for CMB science, these variables disentangle scalar primordial fluctuations from vectorial and tensorial ones~\citep{Kamionkowski1997}. Indeed, vectorial and tensorial fluctuations in the primordial plasma before the decoupling would give rise to a non-zero $B$-mode signal in the CMB, while scalar fluctuations cannot source any $B$-mode signal. Second, contrary to the complex $Q+iU$ variable which is a spin-2 variable, $E$ and $B$ variables are scalar and pseudo-scalar\footnote{Contrary to scalar quantities, the sign of pseudo-scalar quantities is changed under a parity inversion.} variables, respectively, which means in particular that these variables are invariant under any rotation of the initial basis chosen for the measurement of the polarization signal. This explains why the choice of the axis with respect to which the $\phi$ angle introduced in Eqs.~\eqref{eq:Q_dust} and \eqref{eq:U_dust} is defined has no impact on these variables.

Following the procedure explained in this section, we thus compute $(I_\nu, E_\nu, B_\nu)$ maps for five different frequency channels: $\nu = 300, 500, 800, 1500, 3000$~GHz. This choice of channels is inspired by the analysis of \cite{planck2013-p06b} on \textit{Planck} and IRAS data over the 353-3000~GHz range. For this range of frequencies, the effects of $\beta$ and $T$ are expected to be disentangled. As an example, we display in Fig.~\ref{fig:data_IEB} the resulting maps $I_\nu$, $E_\nu$ and $B_\nu$ (from left to right, respectively) for the three following channels: 300, 800, and 3000~GHz. The filamentary structure of these maps underlines the non-Gaussianity of the data. Moreover, at a given frequency, coherent structures visible on the total intensity map $I_\nu$ have obvious counterparts in the polarization maps $E_\nu$ and $B_\nu$. Statistical correlations from one observable to another are thus expected. Across the frequency axis, maps of the same observable seem to be almost structurally identical. Except for the important variation of the standard deviation of the maps, and the variation of the means for total intensity maps, the impact of frequency on the general aspect of the maps appears to be very subtle. To better see these subtle variations, we show in Fig.~\ref{fig:synth_I_multifreq_maps} (top row, center and right) examples of total intensity ratio maps between consecutive frequency channels.

%% file: methodo.tex
\section{Formalism}
\label{sec:methodology}

This section introduces the formalism underlying the definition of our generative models, namely microcanonical gradient descent models conditioned by WPH statistics. Samples of microcanonical gradient descent models are drawn by iteratively deforming an initial Gaussian sample into a signal with some prescribed statistics. Here, we choose these prescribed statistics to be WPH statistics, as these have been proven efficient in characterizing coherent structures in a variety of non-Gaussian random fields~\citep{Mallat2020, Allys2020, Zhang2021, Regaldo2021, Jeffrey2021, Brochard2022}.

In Sect.~\ref{sec:microcanonical_models}, we briefly define microcanonical gradient descent models. Then, in Sect.~\ref{sec:wph}, we introduce the families of moments on which the models of this paper rely. This concise introduction is complemented by Appendix~\ref{app:app_wph}, which contains all the technical details needed to reproduce this work.

\subsection{Microcanonical gradient descent models}
\label{sec:microcanonical_models}

Microcanonical gradient descent models were introduced in~\cite{Bruna2019} as approximations of maximum entropy microcanonical models.\footnote{We refer the reader to Appendix~\ref{app:microcanonical} for formal definitions of maximum entropy microcanonical models.} With $x$ the realization of the random field $X$ that we want to approximate, and $\phi(x)$ the vector of statistics that is supposed to characterize the statistical properties of $X$, these models are defined by transporting an initial Gaussian distribution through gradient descent over:
\begin{equation}
\label{eq:microcanonical_loss}
\mathcal{L}(y) = \norm{\phi(y) - \phi(x)}^2,
\end{equation}
where $\norm{\cdot}$ is the Euclidean norm.
The gradient descent algorithm defines at each iteration $k$ a mapping $f_k(y) = y - \kappa_k\nabla\mathcal{L}(y)$ with $\nabla\mathcal{L}$ the gradient of $\mathcal{L}$ and $\kappa_k$ the gradient step at iteration $k$. For a given number of iterations $n$, and with $y_0$ a sample drawn from our initial Gaussian distribution, the resulting sample $y_n$ of our model reads $y_n = f_n \circ f_{n-1} \circ \dots \circ f_1(y_0)$. In practice, the number of iterations is empirically adapted to reach an approximate convergence.

\subsection{Wavelet Phase Harmonic statistics}
\label{sec:wph}

Originally introduced in \cite{Mallat2020}, the WPH statistics have been proven successful in characterizing complex coherent structures arising from a variety of two-dimensional non-Gaussian physical fields. In \cite{Zhang2021}, they have been applied to turbulent vorticity fields, in \cite{Allys2020} to density fields of the large-scale structure of the Universe, in \cite{Regaldo2021} and \cite{Jeffrey2021} to \textit{Planck} observations and simulated maps of dust polarization data. In all of these fields, structures stem from highly nonlinear physics. One of the main assets of the WPH statistics is to characterize the resulting non-Gaussianity through an efficient quantification of interactions between scales.

The WPH statistics rely on a set of bandpass and lowpass filters allowing to locally decompose the spectral content of the input maps onto a tiling of Fourier space. These filters are respectively wavelets and Gaussian filters. In this section, we introduce \textit{auto-WPH moments}, defined as covariances of nonlinear transformations of the wavelet transform of a given random field $X$. The nonlinear transformation is a pointwise operation called \textit{phase harmonic}, which is defined below. Then, we extend the definition of such moments to the case of a pair of random fields $(X, Y)$, introducing \textit{cross-WPH moments} designed to quantify interactions between scales across $X$ and $Y$. Finally, we introduce a new family of \textit{auto/cross-scaling moments}, defined in a similar fashion as the auto/cross-WPH moments but involving Gaussian filters instead of wavelets. These are designed to better constrain the large-scale behavior of the input fields as well as their (joint) one-point distribution. The WPH statistics of a map $x$, or a pair of maps $(x, y)$, refer to the estimates of the auto/cross-WPH moments and auto/cross-scaling moments.

\subsubsection{Wavelet transform}
\label{sec:wavelet_transform}

A wavelet is a spatially localized waveform with a zero mean which acts as a bandpass filter. From an initial wavelet $\psi$, called the \textit{mother wavelet}, we build a bank of wavelets $\{\psi_{j, \theta}\}_{j, \theta}$ by dilation and rotation of $\psi$, where $j$ is an index of dilation and $\theta$ is an angle of rotation. Formally, we have:
\begin{equation}
	\psi_{j, \theta}(\ve{r}) = 2^{-2j}\psi(2^{-j}R_{\theta}^{-1}\ve{r}).
\end{equation}
The number of dilations and rotations considered are $J$ and $L$, respectively, so that ${0 \leq j \leq J - 1}$ and ${\theta\in\{k\pi/L, 0\leq k \leq L - 1\} }$. Consequently, our bank of wavelets is made of $J\times L$ wavelets. In the following, we make use of bump-steerable wavelets. These are complex-valued wavelets defined in Appendix~\ref{app:app_filters}.

The wavelet transform of $X$ is finally defined as the set of bandpass-filtered maps $\{ X \ast \psi_{j, \theta}\}_{j, \theta}$, where $\ast$ denotes the convolution operation. These convolutions correspond to local bandpass filterings of $X$ at spatial frequencies centered on modes of the form $\ve{k}_{j,\theta} = k_02^{-j}\ve{u}_\theta$, with $\ve{u}_\theta=\cos\left(\theta\right)\ve{e}_x + \sin\left(\theta\right)\ve{e}_y$ and $k_0$ the central frequency of the mother wavelet $\psi$.

In this study, we work with $512\times 512$ maps, and choose ${J = 8}$ and ${L = 4}$. We show in Fig.~\ref{fig:bump_wavelet} one wavelet from our bank.

\subsubsection{WPH moments}
\label{sec:wph_auto}

\paragraph{Auto-WPH moments}

The auto-WPH moments of $X$ are covariances of the phase harmonics of the wavelet transform of $X$, i.e. these are defined as:
\begin{equation}
C_{\lambda,p,\lambda^\prime,p^\prime}(\ve{\tau}) = \Cov{\left[X \ast \psi_\lambda(\ve{r})\right]^{p}}{\left[X \ast \psi_{\lambda^\prime}(\ve{r} + \ve{\tau})\right]^{p^\prime}},
\end{equation}
with $\lambda$ and $\lambda^\prime$ referring to two oriented scales $(j, \theta)$ and $(j^\prime, \theta^\prime)$, and the bracket denoting the pointwise phase harmonic operator $z \mapsto \left[z \right]^p = |z| \cdot \text{e}^{i p~\text{arg}(z)}$.\footnote{We recall that $\Cov{X}{Y} = \E{\left(X - \E{X}\right)\left(Y - \E{Y}\right)^*}$.}$^,$\footnote{These moments do not depend on the $\ve{r}$ variable because of the homogeneity of $X$.} When applied to a complex $z$, the phase harmonic operator preserves the modulus of $z$ but multiplies its phase by a factor $p$. Note that, for $p=0$, this operation simply consists in taking the modulus of $z$, and for $p=1$ it is the identity. This operator plays a key role to capture efficiently interactions between different scales in $X$. For $\lambda = \lambda^\prime$ and $p = p^\prime = 1$, the corresponding moments are averages of the power spectrum over the bandpass of $\psi_\lambda$, so that this class of moments does include the power spectrum information. We refer to Appendix~\ref{app:wph_properties} for additional details on these moments.

\paragraph{Cross-WPH moments}

The previous moments can be extended to the characterization of interactions between the scales of two different fields $X$ and $Y$. We define such cross-WPH moments as follows:
\begin{equation}
C_{\lambda,p,\lambda^\prime,p^\prime}^\times(\ve{\tau}) = \Cov{\left[X \ast \psi_\lambda(\ve{r})\right]^{p}}{\left[Y \ast \psi_{\lambda^\prime}(\ve{r} + \ve{\tau})\right]^{p^\prime}}.
\end{equation}
Just like before, for $\lambda = \lambda^\prime$ and $p = p^\prime = 1$, the corresponding moments are averages of the cross spectrum over the bandpass of $\psi_\lambda$, so that the cross spectrum information is also captured by this class of moments. Note that, in the field of texture synthesis, a similar class of cross moments has been used to characterize color channels interactions in RGB images~\citep{Vacher2021, Brochard2022}.

\paragraph{Discretization of $\tau$}

The previous moments all depend on the relative shift $\ve{\tau}$ between $X\ast \psi_\lambda$ and $X\ast \psi_{\lambda^\prime}$ (or $Y\ast \psi_{\lambda^\prime}$). Inspired by \cite{Brochard2022}, we discretize this variable as follows:
\begin{equation}
\label{eq:tau_n_alpha}
    \ve{\tau}_{n, \alpha} =
    \begin{cases}
    2^n\ve{u}_{\alpha} &\text{for $n \geq 1$},\\
    \ve{0} &\text{for $n = 0$},
    \end{cases}
\end{equation}
with $0 \leq n \leq \Delta_n - 1$ and ${\alpha\in \{\alpha_k = \frac{k\pi}{A}, 0\leq k \leq 2A-1\}}$, where $\Delta_n$ and $A$ are two integers playing similar roles to $J$ and $L$ but for $n$ and $\alpha$ variables instead of $j$ and $\theta$, respectively. In this study, we choose $\Delta_n = 5$ and $A=4$.\footnote{Since we deal with fields with periodic boundary conditions, in order to avoid redundancy in the coefficients, $\Delta_n$ should verify $\Delta_n < \log_2(N/2) + 1$, where $N$ is the number of pixels along the smallest axis of our images. Our choice meets this criterion.}

\paragraph{Choice of a subset of moments}

Estimating the auto/cross-WPH moments for every possible value of $\lambda$, $\lambda^\prime$, $p$, $p^\prime$, and $\ve{\tau}$ is not an option for several reasons. Besides the fact, that this would be computationally expensive, some moments may vanish or be redundant by construction. Moreover, the number of moments should be kept sufficiently low compared to the dimension of the data to allow for statistical diversity in the resulting generative model. Indeed, when based on a single realization $x$, if the number of statistical constraints is too high, samples of such models will tend to reproduce the specific features of $x$ instead of being representative of the variability of $X$~\citep[for a discussion, see][]{Brochard2022}.

Therefore, we need to choose a reduced subset of moments that will characterize the relevant statistical properties of our data. This is a crucial and difficult modeling step. Although this subset should be dependent on the nature of the data, here our choice is mostly inspired by the literature on this subject~\citep{Allys2020, Regaldo2021, Brochard2022}, intending for it to be relevant for a reasonable variety of non-Gaussian random fields.  We refer to Appendix~\ref{app:wph_subset} for a detailed presentation of this subset.

For our choice of $J$, $L$, $\Delta_n$, and $A$ values, the resulting number of auto-WPH coefficients is 6940, which amounts to $2.6~\%$ of the number of pixels of a $512\times512$ image. However, most of these coefficients are complex-valued numbers, so that the effective dimension of the description is $\sim 1.3 \times 10^4$, leading to a ratio of $\sim 5~\%$. On the other hand, for a pair of maps, we consider 1264 cross-WPH coefficients, with, here again, most of them being complex-valued numbers.

\subsubsection{Scaling moments}
\label{sec:scaling_moments}

To better constrain the large scales as well as the (joint) one-point distribution of the input fields, we introduce a new class of auto/cross-scaling moments relying on a family of isotropic Gaussian filters $\{\varphi_j\}_{j}$. These filters are built by dilating an initial Gaussian function $\varphi$ (defined in Appendix~\ref{app:app_filters}) similarly to what is done in Sect.~\ref{sec:wavelet_transform} for wavelets. Assuming zero-mean $X$ and $Y$, our auto/cross-scaling moments are defined as follows:
\begin{align}
    L_{j, p, p^\prime} &= \Cov{\left[X \ast \varphi_{j}(\ve{r})\right]^{p}}{\left[X \ast \varphi_{j}(\ve{r})\right]^{p^\prime}}, \\
    L_{j, p, p^\prime}^\times &= \Cov{\left[X \ast \varphi_{j}(\ve{r})\right]^{p}}{\left[Y \ast \varphi_{j}(\ve{r})\right]^{p^\prime}}.
\end{align}

In this work, we choose $j \in \{-1, 0, 1, 2\}$ and consider the following set of values for $(p, p^\prime)$: $\{(0, 0), (1, 1), (0, 1)\}$ for auto moments, and $\{(0, 0), (1, 1), (0, 1), (1, 0)\}$ for cross moments.\footnote{Note that this choice of $(p, p^\prime)$ values for the cross moments makes the resulting set of moments invariant under the exchange of $X$ and $Y$.} These moments complement our description with a very small number of coefficients: 12 and 16 in the auto and cross cases, respectively.

\subsubsection{Estimation and computation}

We estimate the previous moments from a given map $x$ or couple of maps $(x, y)$ as explained in Appendix~\ref{app:wph_normalization}. Note that the resulting statistical coefficients include a normalization designed to facilitate the gradient descent involved during the sampling of our model (see Sect.~\ref{sec:microcanonical_models}). We denote by $\phi_{\rm auto}(x)$ and $\phi_{\rm cross}(x, y)$ the corresponding vector of auto and cross-WPH statistics, respectively. Computations all employ the GPU-accelerated Python package \texttt{PyWPH}~\citep{Regaldo2021}.

%% file: models.tex
\section{Models and their validation}
\label{sec:models}

This section presents the core results of this paper: the construction and validation of two distinct generative models built from the simulated data introduced in Sect.~\ref{sec:presentation_data}. These are:
\begin{enumerate}
    \item A mono-frequency model based on the joint observation of $(I_{300}, E_{300}, B_{300})$, introduced in Sect.~\ref{sec:monofreq_model}.
    \item A multi-frequency model based on the joint observation of $(I_{300}, I_{500}, I_{800}, I_{1500}, I_{3000})$, introduced in Sect.~\ref{sec:multifreq_model}.
\end{enumerate}
For each of these models, we perform a visual and quantitative assessment of their realism.

The data and codes to reproduce the models are provided on \url{https://github.com/bregaldo/dust_genmodels}.

\subsection{Mono-frequency model}
\label{sec:monofreq_model}

\subsubsection{Description of the model}

We build a generative model based on the joint observation of $(I_{300}, E_{300}, B_{300})$ as follows (referred to as ``the WPH model" in the following). We define a microcanonical model of $\ve{x} = (\log(I_{300}), E_{300}, B_{300})$ conditioned by the following descriptive statistics:
\begin{align}
    \phi(\ve{x}) &= \phi_{\rm auto}(\log(I_{300})) \oplus \phi_{\rm auto}(E_{300}) \oplus \phi_{\rm auto}(B_{300}) \nonumber \\
    &\oplus \phi_{\rm cross}(\log(I_{300}), E_{300}) \oplus \phi_{\rm cross}(\log(I_{300}), B_{300})),
\end{align}
where $\oplus$ denotes vector concatenation. We choose to model the logarithm of the total intensity map $I_{300}$ for two main reasons: 1) the logarithm tends to Gaussianize the data (analytical statistical models of dust total intensity maps are usually defined as log-normal fields, see e.g. \cite{Levrier2018}), and thus to simplify the modeling, 2) pixels of $I$ maps take positive values, so that working with the logarithm avoids to impose positive values on the synthetic $I$ maps during the sampling procedure. Also note that $\phi_{\rm cross}$ is designed to be \textit{pseudo}-symmetric (see Appendix~\ref{app:cross_wph_subset}), so that $\phi_{\rm cross}(\log(I_{300}), E_{300})$ and $\phi_{\rm cross}(E_{300}, \log(I_{300}))$ are equally informative in characterizing the couplings between $\log(I_{300})$ and $E_{300}$ (and obviously this remains true for $(\log(I_{300}), B_{300})$). Our choice to ignore the couplings between $E_{300}$ and $B_{300}$ is motivated by observational measurements indicating a null $EB$ cross-spectrum~\citep{planck2016-l11A}. However, note that these measurements do not say anything on higher-order cross statistics, so that this modeling choice remains partly arbitrary.

The effective dimension of $\phi(\ve{x})$ is $\sim 4.4\times 10^4$, which amounts to less than 6~\% of the total number of pixels of $\ve{x}$. Samples are drawn by minimizing the objective function defined in Eq.~\eqref{eq:microcanonical_loss}, starting from a triplet of realizations of three independent Gaussian random fields having the same power spectrum as the empirical power spectrum of $\log(I_{300})$, $E_{300}$, and $B_{300}$, respectively.\footnote{Note that the power spectrum of the maps is constrained during the optimization, thus we could have also started from independent realizations of Gaussian white noises without impacting the results of this paper. Here, this choice is only motivated by the numerical efficiency of the optimization.} To perform this optimization we use the L-BFGS algorithm that is implemented in \texttt{SciPy}~\citep{Byrd1995, Zhu1997, Virtanen2020}.\footnote{The L-BFGS algorithm is a quasi-Newton method, which is not, properly speaking, a gradient descent method. However, this algorithm has been shown experimentally to be more suitable for solving this optimization problem compared to standard gradient descent algorithms.} This necessitates the computation of the gradients of the objective with respect to the pixels of the optimized maps at every iteration. We compute such gradients using automatic differentiation as implemented in \texttt{PyTorch}~\citep{Paszke2019}. The sampling takes $\sim 6$~min for 200~iterations on a NVIDIA A100-SXM4 GPU with 40GB of memory.

In the following, we also make use of a Gaussian equivalent of this previous model as a baseline (referred to as ``the Gaussian model"). It is a microcanonical model conditioned by the auto and cross-WPH statistics that only estimate the power and cross spectrum information, respectively. As before, we ignore the couplings between $E_{300}$ and $B_{300}$. We refer to Appendix~\ref{app:app_gaussian} for additional details on this model and to Fig.~\ref{fig:synth_IEB_monofreq_gaussian_maps} for visual examples of its samples. The samples are drawn as above.

\subsubsection{Visual assessment}

\begin{figure*}
    \centering
    \includegraphics[width=0.85\hsize]{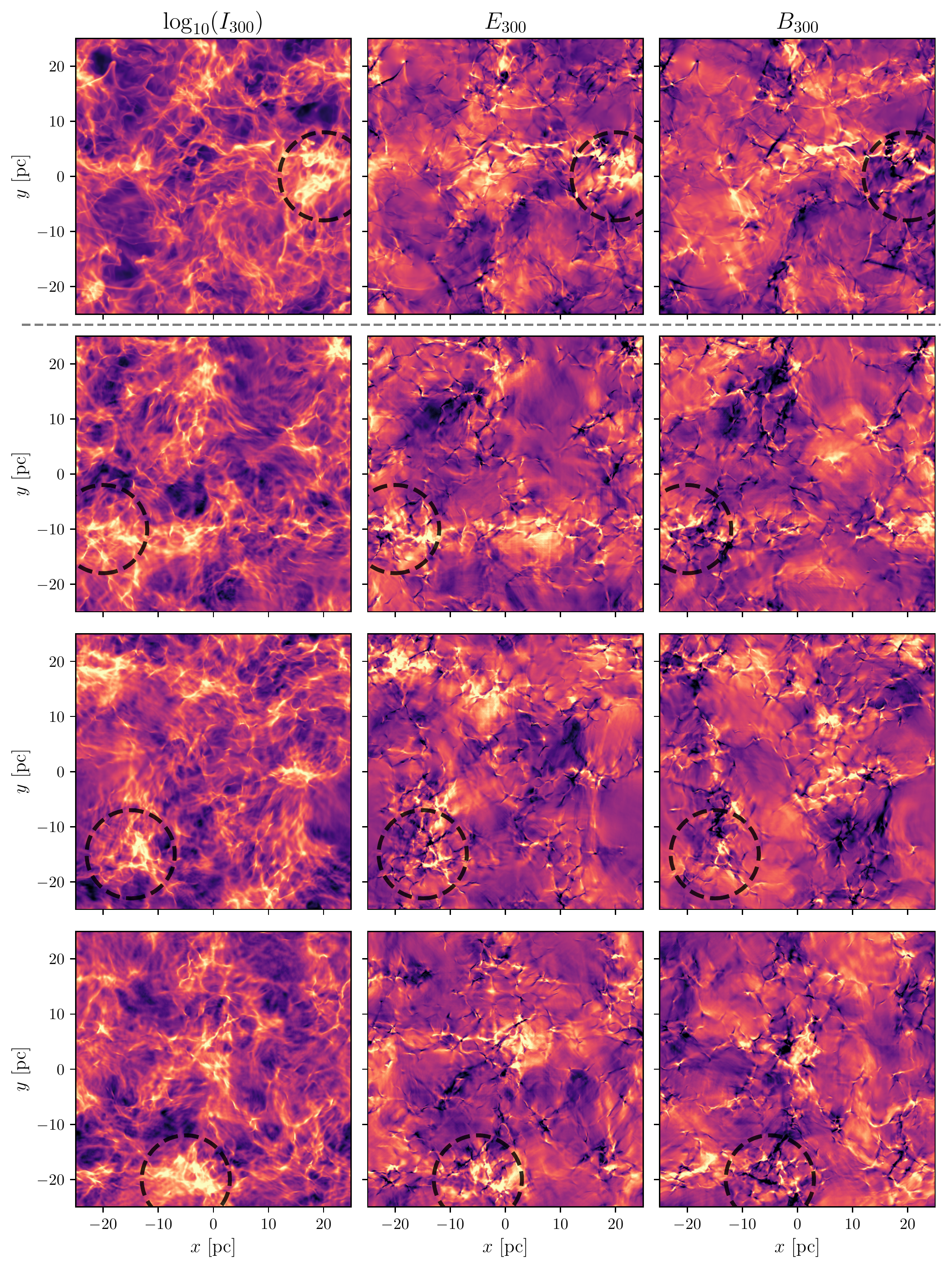}
    \caption{Joint $(\log_{10}(I_{300}), E_{300}, B_{300})$ syntheses (from second to last rows) sampled from the generative model described in Sect.~\ref{sec:monofreq_model} and which only relies on the statistics of the joint map shown in top row (same joint map as the top row of Fig.~\ref{fig:data_IEB}). Dashed circles show examples of spatial correlations across the observables $I$, $E$, and $B$.}
    \label{fig:synth_IEB_monofreq_maps}
\end{figure*}

We show in Fig.~\ref{fig:synth_IEB_monofreq_maps} the original simulated maps $(\log(I_{300}), E_{300}, B_{300})$ (first row, from left to right) next to three different samples (also referred to as ``syntheses") of the WPH model (second to fourth rows).

The synthetic $\log(I_{300})$ maps statistically reproduce the main features of the original map. Given the important variability of structures across samples, and the difficulty to distinguish original from synthetic maps at first glance, this model seems to provide a relevant approximation of the underlying probability distribution of the original data. We emphasize that to achieve similar results, deep generative models usually require thousands of observations, whereas, here, our model is only based on a single example. In \cite{Aylor2021} and \cite{Thorne2021}, slightly more than one thousand total intensity maps were used to train a GAN and a VAE, respectively. While the visual quality of the GAN syntheses is roughly equivalent to ours, this is not the case for that of the VAE model which notably fails to reproduce small-scale patterns of the maps.

For the polarization maps $E_{300}$ and $B_{300}$, there is also qualitative visual agreement between the synthetic and original maps, although close scrutiny reveals subtle artifacts compared to the $I$ case. The diversity of structures across samples is still satisfactory in this case.

Finally, spatial correlations between the total intensity maps and the polarization maps are also well reproduced. Circles in black dashed lines show examples of such correlations for both the original and synthetic data. Taking into account these correlations is a significant improvement over both previous microcanonical models~\citep{Allys2019, Regaldo2020, Jeffrey2021} and deep generative models of dust emission maps~\citep{Aylor2021, Thorne2021, Krachmalnicoff2021}.

\subsubsection{Quantitative assessment}

\begin{figure*}
    \centering
    \includegraphics[width=0.9\hsize]{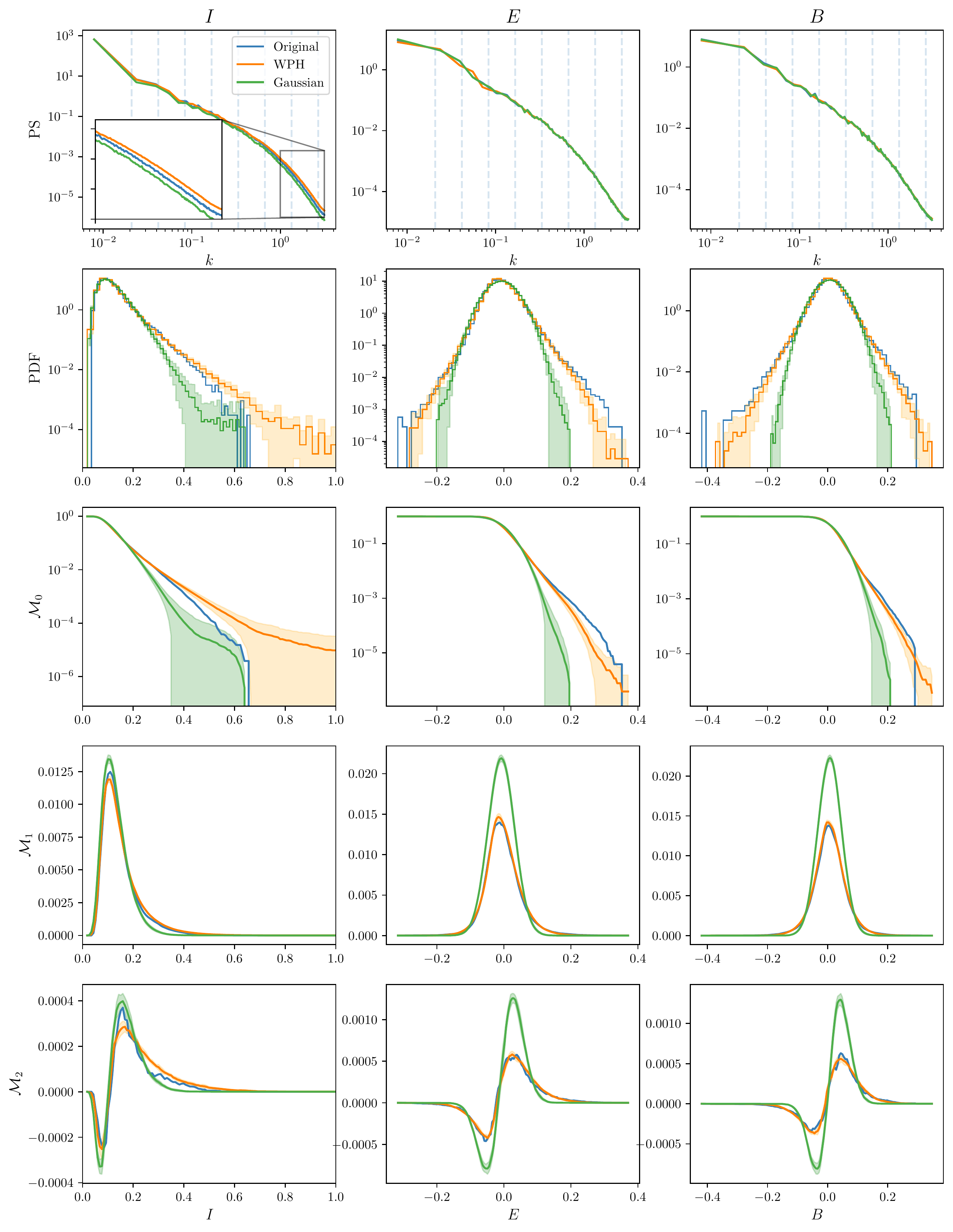}
    \caption{Quantitative validation of the mono-frequency model described in Sect.~\ref{sec:monofreq_model}. We compare the empirical power spectra (first row), distributions of pixel values (second row), and Minkowski functionals (third to fifth rows) computed separately for the $I$, $E$, and $B$ maps (from left to right) of the original, WPH and Gaussian data. For the WPH and Gaussian models, we show mean statistics across ten independent samples, and the errors are the standard deviations across these samples. Vertical dashed lines on the power spectrum plots mark the central frequencies of the wavelets used for this analysis.}
    \label{fig:synth_IEB_monofreq_stats}
\end{figure*}

\begin{figure}
    \centering
    \includegraphics[width=\hsize]{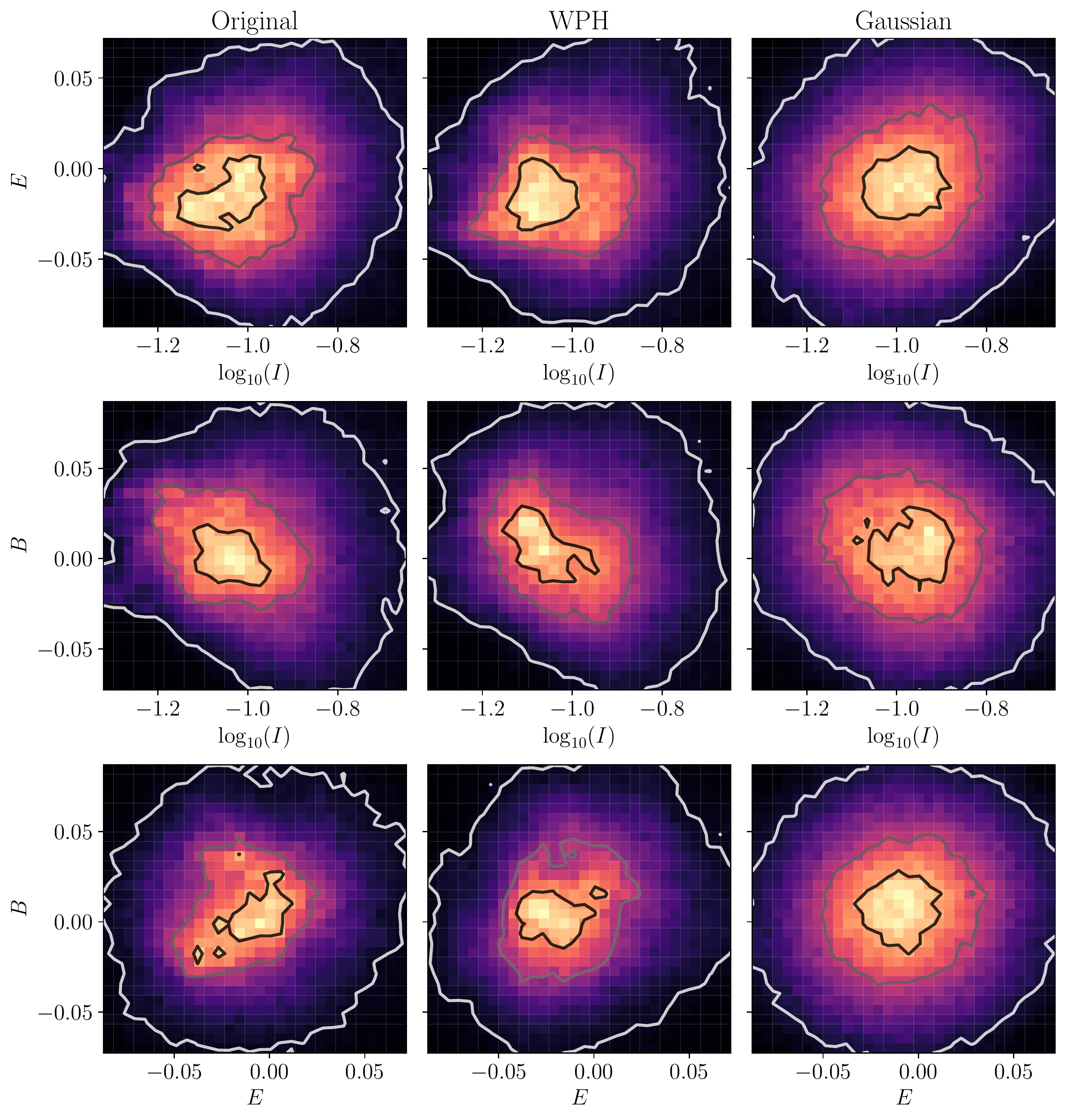}
    \caption{Quantitative validation of the mono-frequency model described in Sect.~\ref{sec:monofreq_model}. We compare the joint distributions of the pixel values of $(\log(I), E)$, $(\log(I), B)$, and $(E, B)$ (from top to bottom) in the original, WPH, and Gaussian maps (from left to right). We add for each of these plots, contours at 0.5$\sigma$, 1$\sigma$, and 2$\sigma$. For the WPH and Gaussian models, statistics are estimated from a single sample.}
    \label{fig:synth_IEB_monofreq_stats_hist2d}
\end{figure}

We choose to quantitatively assess the realism of the WPH model by means of a statistical comparison. We build ten independent samples of both the WPH and the Gaussian model, following the same procedure as before. In Fig.~\ref{fig:synth_IEB_monofreq_stats}, we compare the empirical power spectra, distribution of pixels values, and the Minkowski functionals derived from the samples of the WPH model, those of the Gaussian model, and the original maps, for each of the observables $I$, $E$, and $B$ taken separately. For the WPH Gaussian models, we show mean statistics across the ten independent samples, and when displayed, the error bars correspond to the standard deviations across these samples. In Fig.~\ref{fig:synth_IEB_monofreq_stats_hist2d}, we compare the joint distributions of pixel values between all pairs of observables in the original, WPH, and Gaussian cases. Contrary to Fig.~\ref{fig:synth_IEB_monofreq_stats}, in the WPH and Gaussian cases, we show statistics estimated from a single sample of the model. The results are discussed below.

\paragraph{Power spectrum.}

We first recall that our models directly include power spectrum constraints (see Appendix~\ref{app:wph_properties} for additional details), so that the power spectra of the synthetic maps are expected to match those of the original maps very well. We show the power spectrum analysis in the top row of Fig.~\ref{fig:synth_IEB_monofreq_stats}. While the power spectra of the $E$ and $B$ maps are indeed very well reproduced, the agreement is slightly worse for the $I$ maps at small scales. The discrepancies take the form of a very subtle excess of power in the synthetic maps. As explained previously, we constrain the WPH statistics of $\log(I)$ instead of $I$, so that this kind of discrepancies is not surprising. We have checked that the power spectrum of $\log(I)$ is very well reproduced, as expected. These discrepancies thus show that the WPH model does not perfectly capture the power spectrum of the exponential of the constrained data. Similarly, we observe comparable discrepancies between the power spectrum of the Gaussian data and that of the original data. Here again, the Gaussian model applies to $\log I$ and not $I$, meaning that $I$ is modeled by a log-Gaussian model. Such discrepancies thus underlie the limit of log-Gaussian models for dust intensity maps.

\paragraph{Distribution of pixels.}

We show the distributions of pixel values in the second row of Fig.~\ref{fig:synth_IEB_monofreq_stats}. The bulks of these distributions agree between the syntheses and the original maps to a very good extent (three orders of of magnitude on the $y$-axis). Discrepancies appear in the tails of the distributions. These are subtle for the polarization observables $E$ and $B$, but more significant for the $I$ maps. Note that \cite{Aylor2021} pointed out similar difficulties in capturing the tails of these distributions with their GAN model. The quality of the constraints on these statistics within the WPH model highly depends on our choice of scaling moments. It is likely that we could improve these results by taking into account more $j$ values in the definition of our subset of scaling moments (see Sect.~\ref{sec:scaling_moments}). Compared to the Gaussian model, the WPH model does much better, showing its ability to capture non-Gaussian properties of the data.

\paragraph{Minkowski functionals.}

Minkowski functionals are often used to characterize the morphological aspects of smooth random fields. In cosmology, these have been used in various contexts, as the investigation of potential non-Gausianity and anisotropy in the CMB~(see e.g. \cite{planck2014-XXIII, planck2016-l07}), the characterization of the large-scale structure~(e.g.~\cite{Codis2013}) or that of weak lensing data~(e.g. \cite{Parroni2020}). In the ISM community, these are however much less popular statistics, although it has already been applied in the context of dust modeling~\citep{Aylor2021, Krachmalnicoff2021, Burkhart2021}. In two dimensions, there are three Minkowski functionals $\mathcal{M}_0$, $\mathcal{M}_1$, and $\mathcal{M}_2$, which are defined as follows. For given a map $x$ and $\alpha \in \mathbb{R}$, we define the excursion set $\Gamma_\alpha = \{\ve{r}~\lvert~x(\ve{r}) \geq \alpha\}$, which simply corresponds to the region where the field is greater than a given threshold. The Minkowski functionals associated with $x$ are:
\begin{align}
\mathcal{M}_0(\alpha) &= \frac1{A}\int_{\Gamma_\alpha}\dI a, \\
\mathcal{M}_1(\alpha) &= \frac1{2\pi A}\int_{\partial \Gamma_\alpha}\dI l, \\
\mathcal{M}_2(\alpha) &= \frac1{2\pi^2 A}\int_{\partial \Gamma_\alpha}\kappa\dI l,
\end{align}
where $A$ is the total area of the field, $\partial \Gamma_\alpha$ is the boundary of $\Gamma_\alpha$, $\dI a$ and $\dI l$ are the surface and line elements associated with $\Gamma_\alpha$ and $\partial \Gamma_\alpha$, respectively, and $\kappa$ is the curvature of ${\partial \Gamma_\alpha}$. The functionals $\mathcal{M}_0$, $\mathcal{M}_1$ and $\mathcal{M}_2$ are called the area, perimeter, and genus (i.e. number of ``holes") characteristics, respectively.

We show in the third, fourth, and fifth rows of Fig.~\ref{fig:synth_IEB_monofreq_stats}, the Minkowski functionals statistics for our data. We compute these statistics using the Python package \texttt{QuantImPy}~\citep{Boelens2021}. The area $\mathcal{M}_0$ just gives another perspective of the previous distributions of pixels as it directly relates to the cumulative distribution of pixels values. Contrary to this, the perimeter $\mathcal{M}_1$ and genus characteristics $\mathcal{M}_2$ provide new insights on our models. We see that for the polarization observables $E$ and $B$, these statistics are very well reproduced by the WPH model. In comparison, the Gaussian model performs poorly. Comparable results on polarization data were obtained in \cite{Krachmalnicoff2021} using a GAN (although in a slightly different context). Being able to reproduce such results with a model based on a single observation and without training is one of the successes of our approach. On the other hand, for $I$, although the WPH model seems to do slightly better than the Gaussian model for $\mathcal{M}_1$, this is less clear for $\mathcal{M}_2$. This underlies the higher difficulty of the WPH model to perfectly capture the complexity of $I$ maps. Note that we would get similar results for $\log(I)$ on $\mathcal{M}_1$ and $\mathcal{M}_2$ since applying a pointwise monotonous function to our maps would not impact the level sets involved in the derivation of these statistics.

\paragraph{Joint distribution of pixels.} Figure~\ref{fig:synth_IEB_monofreq_stats_hist2d} shows the joint distributions of pixel values of $(\log(I), E)$, $(\log(I), B)$, and $(E, B)$ (from top to bottom row) for our data. These statistics are, here again, mostly constrained by the scaling moments we have introduced in the WPH model. The agreement of these distributions between the WPH model and the original data is satisfactory, and significantly better than the Gaussian case. Note that no cross constraint between $E$ and $B$  was imposed, so that it is not surprising to find a slightly worse agreement of the contours. In order to make sure that this agreement is not a mere consequence of the agreement of the marginal distributions of pixel values, we have estimated the mutual information associated with each of these distributions.\footnote{The mutual information of two random variables $X$ and $Y$ quantifies the mutual dependence between $X$ and $Y$, and is defined as the Kullback-Leibler divergence of the joint distribution $p_{X, Y}$ from the product of the marginal distributions $p_Xp_Y$.} The results of this analysis are shown in Table~\ref{table:mutual_information}. These show a significant dependence between all pairs of observables on the original data, as well as a significantly better capture of these dependencies in the WPH model compared to the Gaussian model.

\begin{table}
    \def\arraystretch{1.5}
    \centering
    \begin{tabular}{c|ccc}
        \hline
        \hline
        &Original & WPH & Gaussian \\
        \hline
        $(\log_{10}(I), E)$ & 0.092 & 0.095 $\pm$ 0.006 & 0.033 $\pm$ 0.002 \\
        $(\log_{10}(I),B)$ & 0.122 & 0.101 $\pm$ 0.006 & 0.020 $\pm$ 0.003 \\
        $(E,B)$ & 0.059 & 0.054 $\pm$ 0.008 & 0.014 $\pm$ 0.002 \\
        \hline
    \end{tabular}
    \caption{Mutual information (in shannon units) associated with the joint distributions of pixel values of $(\log_{10}(I), E)$, $(\log_{10}(I), B)$, and $(E, B)$, estimated for the original, WPH, and Gaussian data. In the WPH and Gaussian cases, the values correspond to the means and standard deviations of the mutual information across ten independent samples.}
    \label{table:mutual_information}
\end{table}

\subsection{Multi-frequency model}
\label{sec:multifreq_model}

\subsubsection{Description of the model}

We now define a generative model of the multi-frequency simulated maps $(I_{300}, I_{500}, I_{800}, I_{1500}, I_{3000})$ as follows. For the same reasons as before, we prefer to work with the logarithm of these intensity maps. Since the maps from one frequency channel to another look very similar (modulo a global scaling or shift of the mean), we choose to impose statistical constraints on the differences of the logarithms of these maps (which are also the logarithms of the ratios of these maps) between consecutive frequency channels. One of the maps also needs to be constrained in absolute to serve as a reference. Consequently, with ${\nu_1, \dots, \nu_5 = 300, \dots, 3000}$~GHz, we define a microcanonical model of $\ve{x} = (\log(I_{\nu_1}))\oplus(\log(I_{\nu_{i+1}}/I_{\nu_i}))_{1\leq i\leq 4}$ conditioned by:
\begin{equation}
    \phi(\ve{x}) = (\phi_{\rm auto}(x_i))_{1\leq i \leq 5} \oplus (\phi_{\rm cross}(x_i, x_j))_{1\leq i < j \leq 5}.
\end{equation}

Contrary to the mono-frequency case, we consider cross constraints between all the possible pairs of maps. Although this significantly increases the dimensionality of the statistical description as well as the computational cost of the sampling procedure, this gives the best visual and quantitative results in what follows. The effective dimension of $\phi(\ve{x})$ is $\sim 8.9\times 10^4$, which amounts to $\sim 7$~\% of the total number of pixels of $\ve{x}$. The sampling procedure takes $\sim 17$~min for 200~iterations on a NVIDIA A100-SXM4 GPU with 40GB of memory. We make use of a single sample of this model in the following.

\subsubsection{Visual assessment}

\begin{figure*}
    \centering
    \includegraphics[width=\hsize]{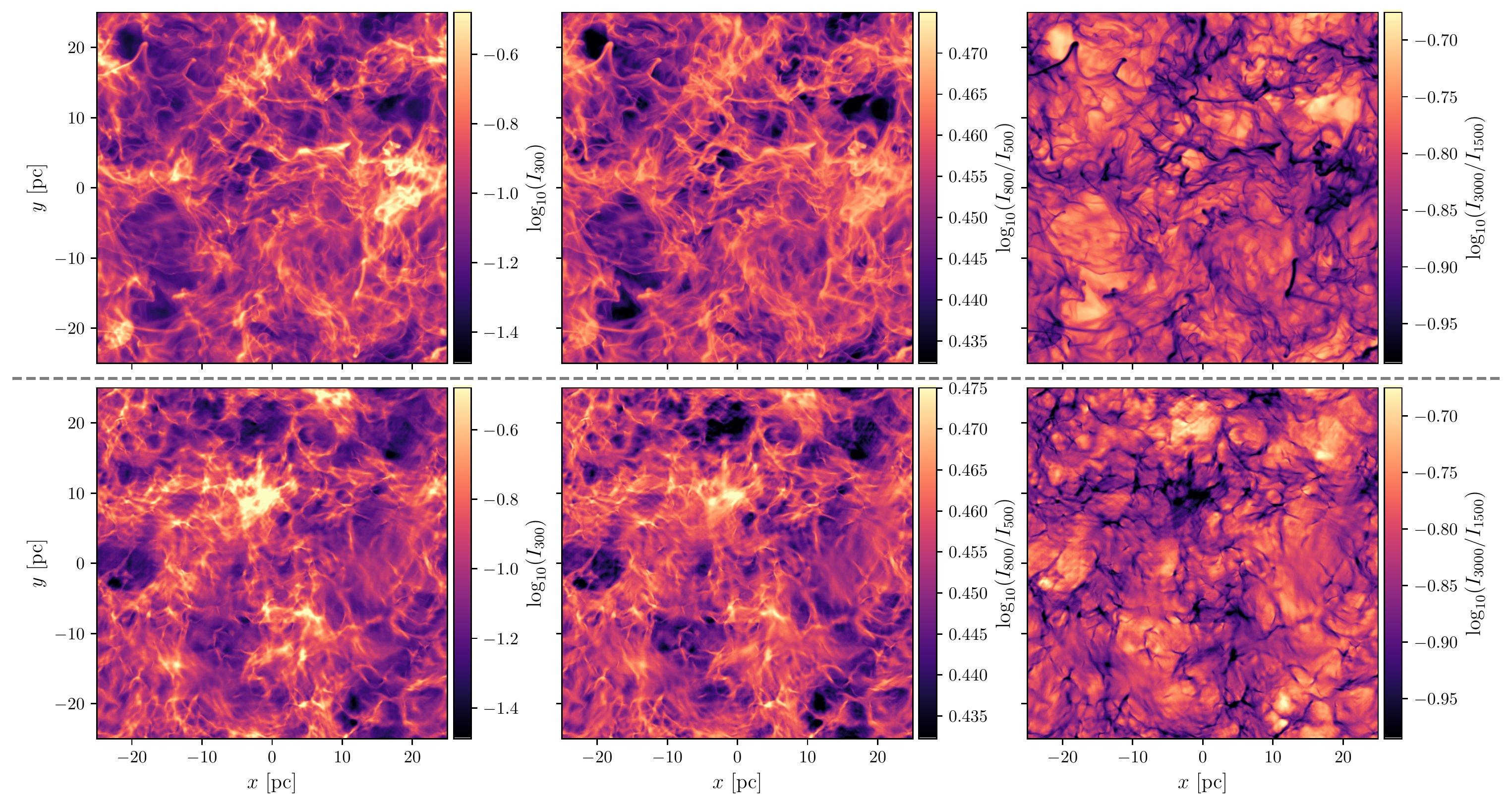}
    \caption{$\log_{10}(I_{300})$, $\log_{10}(I_{800}/I_{500})$, and $\log_{10}(I_{3000}/I_{1500})$ synthetic maps (bottom row, from left to right) sampled from the generative model described in Sect.~\ref{sec:multifreq_model} and which is based on the maps shown in top row.}
    \label{fig:synth_I_multifreq_maps}
\end{figure*}

We show in Fig.~\ref{fig:synth_I_multifreq_maps} a selection of maps derived from the resulting multi-frequency synthesis (bottom row) below the corresponding original maps (top row). We show the $\log(I_{300})$ maps, and the ratio maps $\log(I_{800}/I_{300})$ and $\log(I_{3000}/I_{1500})$ (from left to right). Visually, the synthetic maps are very satisfactory, with a realistic filamentary structure. Moreover, spatial correlations from one map to another are consistent with those visible on the original maps. This shows that our statistical description seems to properly capture interactions between frequency channels.

\subsubsection{Quantitative assessment}

\begin{figure}
    \centering
    \includegraphics[width=\hsize]{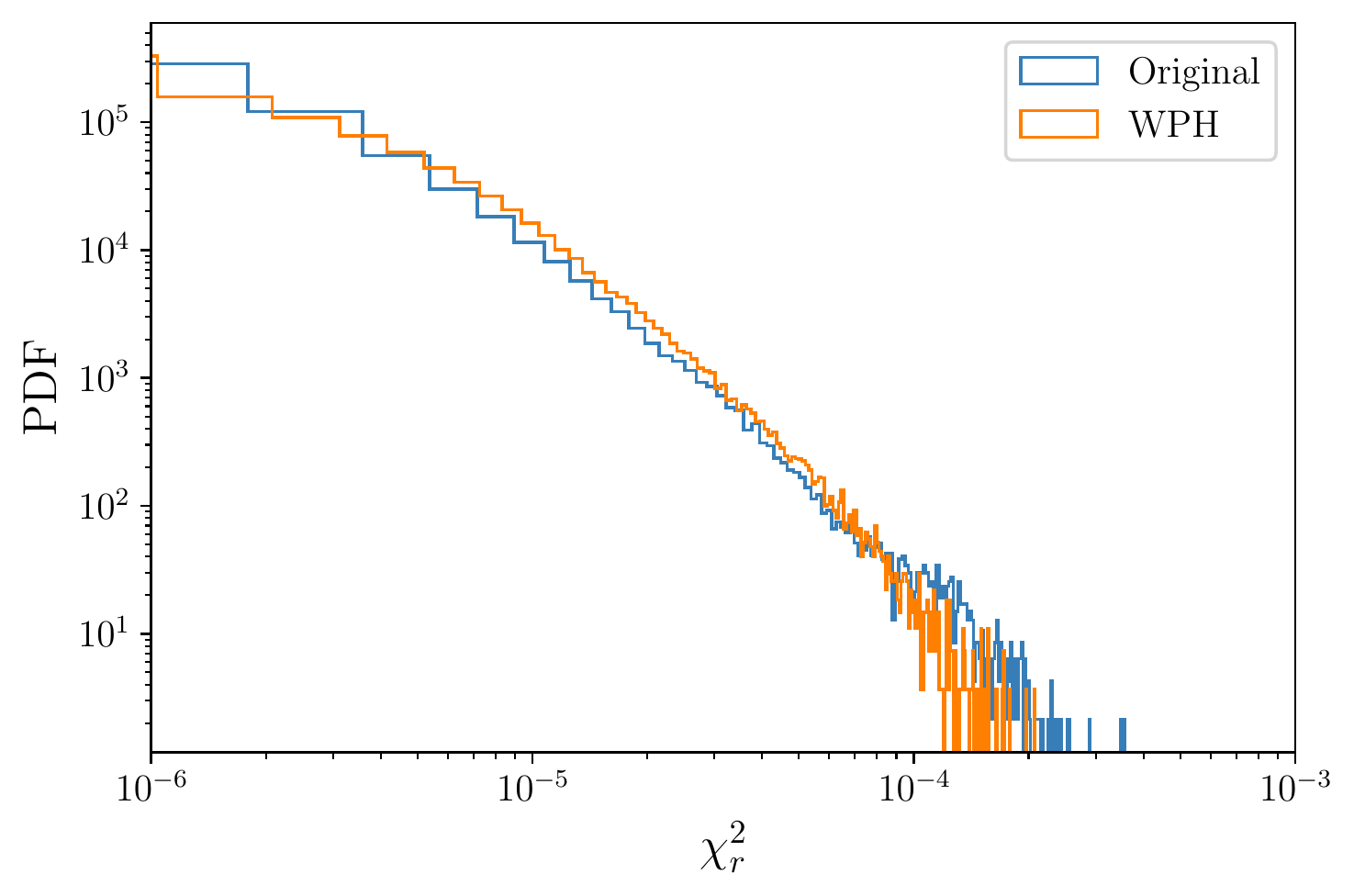}
    \caption{Distributions of $\chi^2_r$ values associated with the pixelwise MBB fits on the original and synthetic multi-frequency $I$ maps.}
    \label{fig:synth_I_multifreq_mbb_chi2r}
\end{figure}

\begin{figure*}
\gridline{\fig{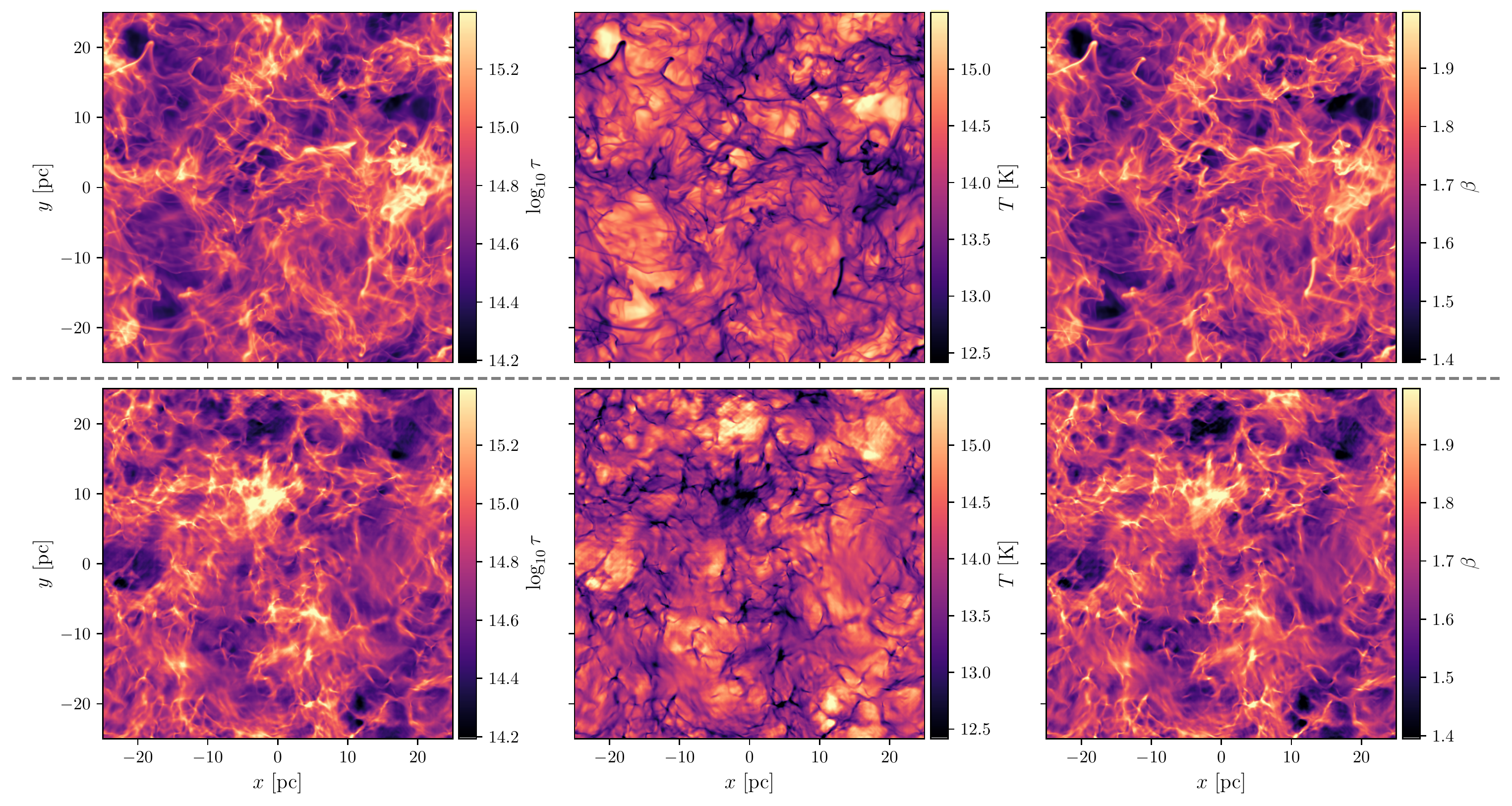}{1.0\textwidth}{(a)}}
\gridline{\fig{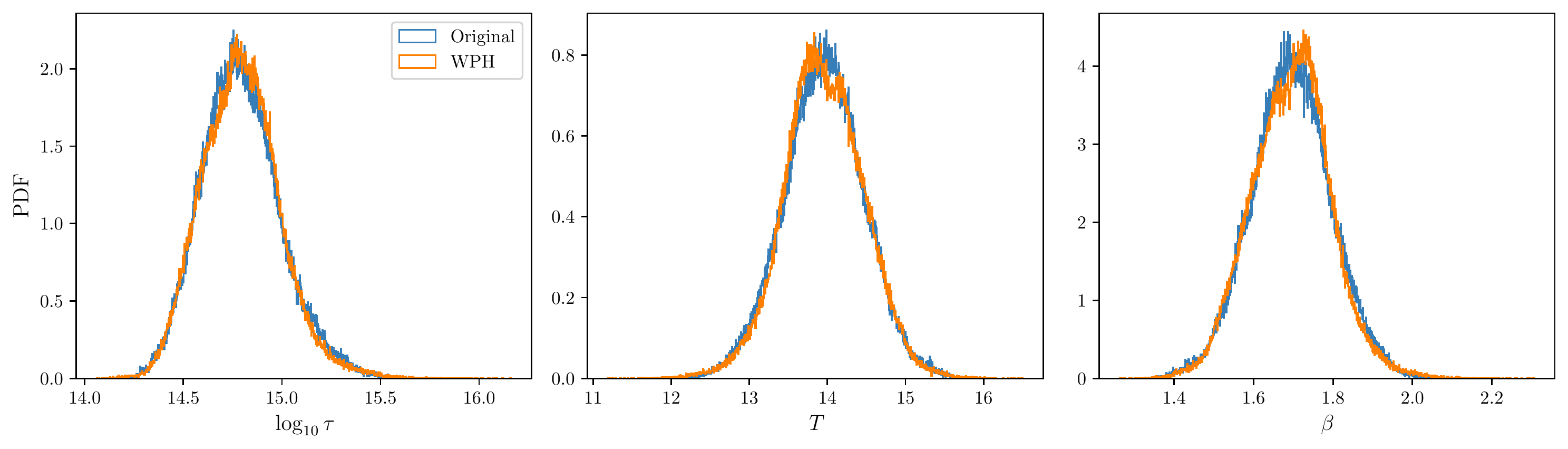}{0.9\textwidth}{(b)}}
\caption{(a) MBB parameters maps $\log_{10}(\tau)$, $T$, and $\beta$ (from left to right) resulting from the pixelwise fits of a MBB model to both the original (top row) and synthetic (bottom row) multi-frequency $I$ maps. (b) Resulting distributions of these parameters.}
    \label{fig:synth_I_multifreq_stats}
\end{figure*}

We quantitatively assess the realism of such multi-frequency models by focusing on how well both the original and synthetic multi-frequency maps can be fitted by a MBB model, and comparing their respective parameters. MBB models are ubiquitous in the interstellar dust literature, and constitute the standard means to model and parameterize the frequency dependence of emission maps (see e.g. \cite{planck2013-p06b}). These are directly related to the model of the emission and absorption properties of dust grains that was employed in Sect.~\ref{sec:presentation_data} to build our original simulated maps. Indeed, a MBB model $\tilde{I}_\nu(\ve{r})$ is defined as:
\begin{equation}
    \tilde{I}_\nu(\ve{r}) = \tau(\ve{r})\nu^{\beta(\ve{r})}B_\nu(T(\ve{r})),
\end{equation}
where, added to the spectral index map $\beta(\ve{r})$ and temperature map $T(\ve{r})$, we also introduce the optical depth map $\tau(\ve{r})$.

For each pixel of both the original and synthetic multi-frequency data, we fit the parameters of this model to the data by performing a non-linear least squares minimization. To do that, we employ the \texttt{SciPy} implementation of the Levenberg-Marquardt algorithm~\citep{Virtanen2020}. We perform the minimization on logarithmic values, and initialize the parameters with $T_0 = 20$~K, $\beta_0=1.5$, and $\tau_0 = I_{300}(\ve{r}) / (300^{\beta_0}B_\nu(T_0))$. We show in Fig.~\ref{fig:synth_I_multifreq_mbb_chi2r}, the resulting distribution of $\chi^2_r$ values. The $\chi^2_r$ values are nowhere larger than $\sim 3\times10^{-4}$, showing that the fit performs very well on both the simulated and synthetic data.\footnote{Since we perform the fit on logarithmic values, $\chi^2_r$ roughly correspond to the mean square relative error.} This result is not particularly surprising for the original data (albeit non-trivial), given the fact that we had employed a MBB-like law of dust grains emission at the voxel level. However, no prior information on this MBB law was given in our model, which shows that our statistical description has been somehow able to capture this SED. Moreover, the agreement of the $\chi^2_r$ distributions is a stronger result as it shows that the MBB is as well suited to model the SED of the synthetic maps as it is for the original maps, which is the ideal behavior of such a model.

We show in Fig.~\ref{fig:synth_I_multifreq_stats} (a) the resulting parameters maps $\tau$, $T$, and $\beta$ for both the original (top row) and synthetic data (bottom row), as well as their distributions (b). Let us first remark that the anti-correlation between $\beta$ and $T$ as well as the positive correlation between $n$ and $T$ that were instilled at the voxel level when building the simulated data in Sect.~\ref{sec:data_maps} clearly reflects in the parameter maps of this (projected) simulated data. Indeed, $T$ clearly tends to be higher (lower) for low (high) values of $\tau$, and similarly, $T$ and $\beta$ appear to be anti-correlated. The synthetic parameters maps exhibit the same properties. Moreover, the consistency of the structures between the synthetic and original parameter maps show that our model is able to capture the spatial variations of this frequency dependence in a very satisfactory way. Comparisons of the distributions of these parameters maps strengthens us in this conclusion.

Note that, in real observational conditions, departures from the MBB model are expected in the dust signal. Formalisms to deal with them have been studied in the literature for both total intensity~\citep[see e.g.][]{Chluba2017, Hensley2018, Mangilli2021} and polarization data~\citep{Ichiki2019, Vacher2022, McBride2022}, and we expect such more refined models to play a central role for high-precision dust foreground modeling in the context of primordial $B$-modes detection. In this context, we believe that learning such models directly from the data, as shown here, constitutes a valuable complementary approach to the accurate modeling of this frequency dependence.

%% file: conclusion.tex
\section{Conclusion}
\label{sec:conclusion}

We have introduced a generic methodology using WPH statistics to build generative models of multi-channel non-Gaussian data. Our approach is purely data-driven, and the models can be derived from a single example of simulated or observational data. Models are microcanonical gradient descent models conditioned by the set of WPH statistics of the target data. We have reviewed the main ideas of the underlying formalism and introduced an extended family of WPH statistics enhancing previous works.

In order to characterize the couplings between two different channels, we have introduced the cross-WPH moments. These are key to this paper. They correspond to direct extensions of the (auto-)WPH moments previously introduced in the literature. They include the cross spectrum information, but also quantify non-Gaussian interactions across channels. We also have introduced a new family of moments, called auto/cross-\textit{scaling moments}, that is designed to probe the largest scales and better constrain the single map or joint one-point distribution. The Python package~\texttt{PyWPH}\footnote{\url{https://github.com/bregaldo/pywph/}}, previously released in \cite{Regaldo2021}, has been updated to easily build such multi-channel generative models from two-dimensional data.

Here, we have applied this methodology to multi-channel simulated maps of the interstellar dust emission. The simulated maps were built from a MHD simulation assuming a simple phenomenological model of the multi-frequency emission of dust grains. We have constructed and studied two separate generative models:
\begin{enumerate}
    \item A model of a mono-frequency $(I, E, B)$ input taking into account $I$-$E$ and $I$-$B$ correlations. Samples of this model were shown to exhibit statistically consistent features with those of the original maps, including correlated structures across $I$, $E$, and $B$. A quantitative assessment demonstrated that the empirical power spectra, distributions of pixel values, and Minkowski functionals of the $I$, $E$, and $B$ synthetic maps match the corresponding statistics of the original maps to a good extent. However, slight discrepancies of these statistics were found for the $I$ maps. Finally, an analysis of the joint distributions of pixel values has shown that our model captures the dependencies between the observables to a good extent.
    \item A model of a multi-frequency $(I_\nu)_\nu$ input, for five frequency channels, and taking into account the strong correlations of the maps across the frequency axis. Samples of the model exhibit consistent features and correlations with those of the original maps. We have quantitatively assessed this model by fitting the SED of our data with a MBB law. This law fits very well and consistently both the synthetic maps and the original maps. The comparison of the parameters maps has shown a very good agreement, underlining the success of our approach in capturing the spatial variations of the SED from the data.
\end{enumerate}

The perspectives of this work are numerous.

First, we underline that the statistical validation of our models could be improved. Indeed, a more rigorous approach would be to compare the statistics of the samples of our model to those of a large number of independent samples of the random field $X$ we wish to approximate. We have chosen here to build our model from a single observation $x$, implicitly assuming that the statistics of $x$ are representative of those of $X$ (which relates to an ``ergodic" assumption, see~\cite{Bruna2019}), and consequently, the validation of our models were performed relatively to the statistics of $x$ only. We could go beyond this assumption by studying the variability of the statistics of a collection of samples $\{x_i\}$ of $X$ in our analysis, and validating our model with respect to this variability. However, let us mention that in our case the data relies on a costly MHD simulation, which prevents the construction of an arbitrarily large number of samples~\citep[for a discussion, see][]{Regaldo2020}.

We will address the modeling of multi-frequency polarization maps in a follow-up paper. In the case of multi-frequency $I$ maps, we have constructed a microcanonical model based on the ratio maps between consecutive channels. Proceeding similarly for $E$ and $B$ is not an option since these variables can exhibit null values. Moreover, the nature of the linear polarization observable is vectorial, which might require the introduction of a complex variable $E+iB$, or $Q+iU$ as in \cite{Regaldo2020, Regaldo2021}.\footnote{Note that this is not an obstacle to our methodology, and that the computation of WPH statistics on complex-valued maps is supported by \texttt{PyWPH}.} These modeling choices should take inspiration from the literature on analytical models of the dust polarization SED~\citep{Ichiki2019, Vacher2022, McBride2022}.

From a more general perspective, the cross-WPH statistics open a new way to analyze and combine multi-channel data, allowing us to efficiently describe and model non-Gaussian correlations across different maps. Moreover, given a WPH model of the interactions between two channels $A$ and $B$, and combined with the formalism of microcanonical models, these statistics should allow for statistically relevant predictions of $B$ maps based on the observation of $A$ maps. Such ``conditioned" models will be explored in further works.

In the context of dust modeling for CMB analysis, such generative models are expected to improve forward models of the CMB sky. In~\cite{Jeffrey2021}, a WPH model of dust polarization maps played a crucial role for CMB foreground marginalization in a likelihood-free inference framework. In this work, the inference was performed on mono-frequency maps, the present paper now paves the way to an extension to the multi-frequency case.

Finally, this work is expected to improve statistical denoising methods of dust polarization maps, such as introduced in~\cite{Regaldo2021}. Taking into account the multi-channel aspect of the data should provide more accurate estimations of the statistics of the noise-free emission since observations at different frequency bands are usually affected by independent noise processes. Note that very recently, and in parallel to this work, a significant step has been taken in this direction by \cite{Delouis2022} employing WST statistics. In this paper, the authors extended the WST to the sphere and introduced cross-WST statistics to characterize correlations across observables in order to statistically denoise \textit{Planck} all-sky maps of the dust emission.

%% file: acknowledgements.tex
\section*{Acknowledgments}
We thank Michael O'Brien and Blakesley Burkhart for their help in investigating a bispectrum validation of the models of this paper.

\software{NumPy \citep{Harris2020},
          PyTorch \citep{Paszke2019},
          PyWPH \citep{Regaldo2021},
          QuantImPy \citep{Boelens2021},
          SciPy \citep{Virtanen2020}
          }

%% file: appendices.tex
\begin{appendix}

\section{Maximum entropy microcanonical models}
\label{app:microcanonical}

In this Appendix, we define maximum entropy microcanonical models, which underlie the microcanonical gradient descent models introduced in Sect.~\ref{sec:microcanonical_models}. This presentation is based on \cite{Bruna2019}, and we refer to this paper for additional details.

Microcanonical models are guided by the principle of maximum entropy, which states that the probability distribution that best represents our knowledge of some system is that with the largest entropy (in the sense of information theory).

Let us consider a random field $X$ and $x$ one of its realizations. We want to approximate the distribution of $X$ based on this single realization $x$. In practice, we make some statistical measurements on $x$ that define a vector of statistics $\phi(x)$. We believe $\phi(x)$ to be sufficiently ``exhaustive" to describe the statistical properties of $X$.\footnote{For the statistics of a realization $x$ to be representative of the statistics of $X$, we need to make an additional assumption of ergodicity (see~\cite{Bruna2019}).} Typically, the realization $x$ lives in $\mathbb{R}^m$, while $\phi(x)$ lives in $\mathbb{R}^n$ with $n < m$. We introduce \textit{microcanonical sets} as ensembles of vectors of $\mathbb{R}^m$ whose statistics are ``sufficiently close" to those of $x$. Formally, for a given $\epsilon > 0$, we define the microcanonical set $\Omega_{\epsilon}$ as:
\begin{equation}
\Omega_{\epsilon} = \{y \in \mathbb{R}^m / \norm{\phi(y) - \phi(x)} \leq \epsilon\},
\end{equation}
where $\norm{\cdot}$ is the Euclidean norm on the statistical space $\mathbb{R}^n$.

In this context, a maximum entropy microcanonical model defined on $\Omega_{\epsilon}$ is a probability distribution $\mu_\epsilon$ supported in $\Omega_{\epsilon}$ with maximal entropy. The entropy of a probability distribution $\mu$ here refers to its differential entropy, called $H(\mu)$ and defined as:
\begin{equation}
H(\mu) = -\int f_\mu(y)\log f_\mu(y) \dI y,
\end{equation}
where $f_\mu$ is the probability density function (PDF) associated with $\mu$. Assuming that the function $\phi$ allows $\Omega_\epsilon$ to be compact, this maximum entropy distribution $\mu_\epsilon$ is simply the uniform distribution on $\Omega_\epsilon$ and is defined by its uniform density:
\begin{equation}
f_{\mu_\epsilon}(y) = \frac{1_{\Omega_\epsilon}(y)}{\int_{\Omega_\epsilon}\dI y}.
\end{equation}

The relevance of this kind of models directly depends on our choice of statistical measurements, represented by the function $\phi$, as well as on the value of $\epsilon$ which is a proxy of the volume of the microcanonical set $\Omega_\epsilon$. Ideally, we want to choose $\phi$ and $\epsilon$ so that typical samples of $X$ are contained in $\Omega_\epsilon$, and conversely, typical samples of $\mu_\epsilon$ are representative of those of $X$.

Even if we manage to define relevant $\phi$ and $\epsilon$ so that the corresponding maximum entropy microcanonical model correctly approximates the distribution of $X$ on paper, we still need to find an efficient way to draw samples from $\mu_\epsilon$. Usual strategies make use of Markov chain Monte Carlo algorithms, however these algorithms reach computational limits when the dimension of the samples $m$ increases.\footnote{In general, Markov chain mixing time depends on the exponential of $m$~\citep{Levin2017}.} For the applications of this paper, these are not an option. Microcanonical gradient descent models as defined in Sect.~\ref{sec:microcanonical_models} allow to circumvent this sampling problem, although these are no longer of maximum entropy in general.

\section{Additional details on the WPH statistics}
\label{app:app_wph}

\subsection{Filters}
\label{app:app_filters}

\begin{figure*}
    \centering
    \includegraphics[width=0.9\hsize]{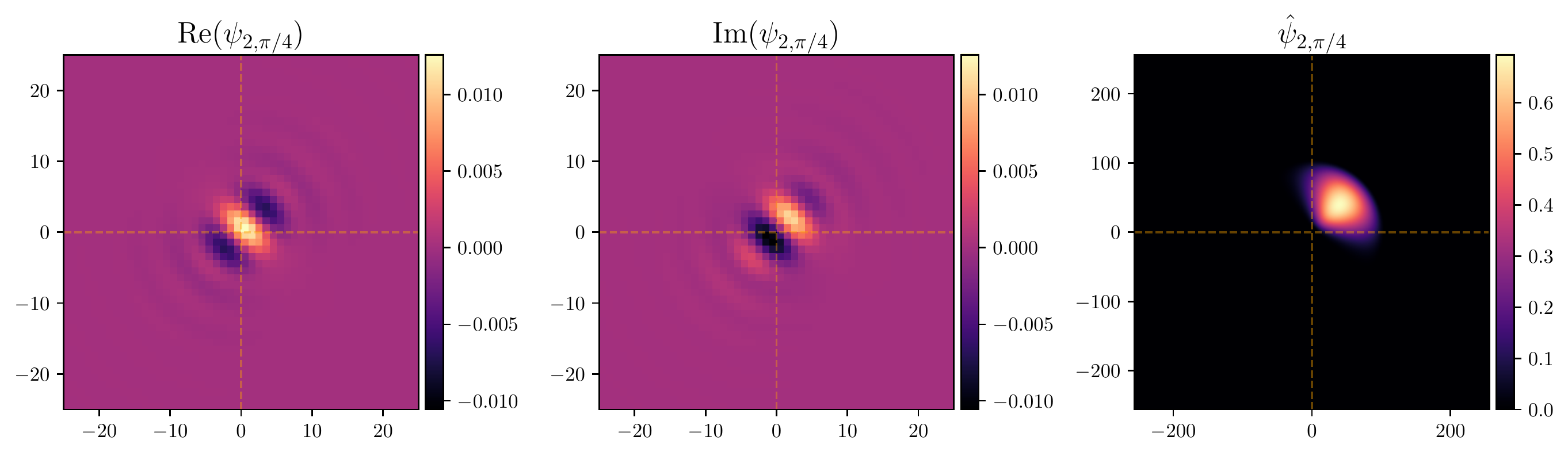}
    \caption{Bump-steerable wavelet $\psi_{2, \pi/4}$ shown in physical space (real part on the left and imaginary part in the middle) and Fourier space (right).}
    \label{fig:bump_wavelet}
\end{figure*}

\paragraph{Bump-steerable mother wavelet} The mother bump-steerable wavelet is defined in Fourier space as follows:
\begin{align}
    \hat{\psi} (\vec k) = &\exp \left(\frac{-(k - k_0)^2}{k_0^2 - (k-k_0)^2}\right) \cdot 1_{[0, 2k_0]}(k) \times \cos^{L-1} (\arg (\vec k)) \cdot 1_{[0, \pi/2]}(|\arg (\vec k)|),
\end{align}
with $k = \norm{\ve{k}}$, $1_A(x)$ the indicator function that returns 1 if $x\in A$ and $0$ otherwise, and $k_0 = 0.85\pi$ the central wavenumber of the mother wavelet. In this paper, we work with $512\times 512$ maps and ${L = 4}$. We show in Fig.~\ref{fig:bump_wavelet} one example wavelet from the resulting bank in both physical and Fourier space.

\paragraph{Gaussian filter}

The Gaussian filters used in this paper are dilated versions of a Gaussian function $\varphi$, which is defined in Fourier space by the following:
\begin{equation}
 \hat \varphi (\ve k) = \exp \left(-\frac{||\ve k||^2}{2 \sigma^2} \right),
\end{equation}
with $\sigma=0.496\times 2^{-0.55} k_0$ (following~\cite{Zhang2021}).

\subsection{Properties of the WPH moments}
\label{app:wph_properties}

The auto-WPH moments are able to capture interactions between different scales of $X$ thanks to the phase harmonic operator. Indeed, the covariance between $X \ast \psi_{\lambda}$ and $X \ast \psi_{\lambda^\prime}$ vanishes when the wavelets $\psi_{\lambda}$ and $\psi_{\lambda^\prime}$ have nonintersecting bandpasses, and it is otherwise a function of the power spectrum of $X$ and of the bandpasses of the wavelets~\citep{Zhang2021, Allys2020}. This is a consequence of the following relation:
\begin{equation}
\label{eq:wph_ps}
C_{\lambda,1,\lambda,1}(\ve{\tau}) = \int S_X(\ve{k})\hat{\psi}_{\lambda}(\ve{k})\conj{\hat{\psi}_{\lambda}}(\ve{k})e^{-i\ve{k}\cdot\ve{\tau}}\dI\ve{k},
\end{equation}
with $S_X$ the power spectrum of $X$~\citep[for a proof, see][]{Zhang2021}. With proper $p$ and $p^\prime$ values, the phase harmonic operator can make $[X \ast \psi_{\lambda}]^{p}$ and $[X \ast \psi_{\lambda^\prime}]^{p^\prime}$ comparable in the sense that they share common spatial frequencies, allowing an extraction of high-order information through their covariance. Conveniently, this operator is a Lipschitz continuous operator with for all $z, z^\prime \in \mathbb{C}^2$, ${|[z]^p - [z^\prime]^p| \leq \text{max}(|p|, 1)|z - z^\prime|}$~\citep{Mallat2020}. This prevents uncontrolled amplifications, and leads to estimators with reduced variance compared to equivalent moments where the phase harmonic operator would be replaced by a standard exponentiation~\citep{Zhang2021}.

\begin{figure}
    \centering
    \includegraphics[width=0.9\hsize]{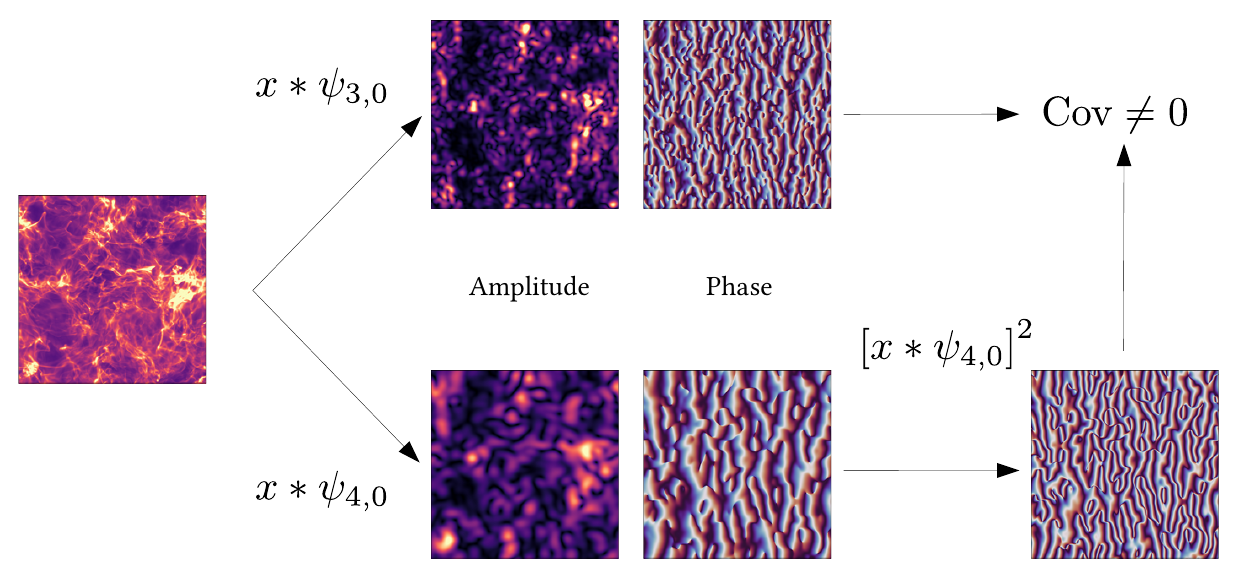}
    \caption{Comparison between the amplitude and phase maps of $x\ast \psi_{j, \theta}$ where $x$ is the map shown on the left and two different wavelets $\psi_{3, 0}$ and $\psi_{4, 0}$. The action of the phase harmonic operator is shown on the phase map of $x\ast\psi_{4, 0}$. Figure inspired by~\cite{Allys2020}.}
    \label{fig:wph_illustration}
\end{figure}

To illustrate the importance of the phase harmonic operator to measure phase alignment between scales, we show in Fig.~\ref{fig:wph_illustration} how the amplitude and phase maps of $x\ast \psi_{\lambda}$ and $x\ast \psi_{\lambda^\prime}$ compare for two different oriented scales $\lambda$ and $\lambda^\prime$, with $x$ the $I_{300}$ map built in Sect.~\ref{sec:data_maps}. We choose $\lambda = (3, 0)$ and $\lambda^\prime = (4, 0)$, so that $\psi_{\lambda}$ and $\psi_{\lambda^\prime}$ probe similar orientations but different scales, with $\psi_{\lambda^\prime}$ probing scales that are twice larger than those probed by $\psi_{\lambda}$. The amplitude maps show local variations of the signal $x$ filtered at different scales, with naturally coarser variations in $|x\ast\psi_{\lambda^\prime}|$ than in $|x\ast\psi_{\lambda}|$. The phase maps present almost periodic oscillations tending to be vertically aligned. However these maps are incoherent in the sense that the average frequency of oscillation of the phase map $\text{arg}(x\ast\psi_{\lambda})$ is approximately twice that of $\text{arg}(x\ast\psi_{\lambda^\prime})$. The phase harmonic operator with $k=2$ transforms $\text{arg}(x\ast\psi_{\lambda^\prime})$ into a phase map (bottom right map) that is much more coherent with $\text{arg}(x\ast\psi_{\lambda})$. Consequently, the sample covariance of these maps does not vanish and quantifies the phase alignment between $x\ast \psi_{\lambda}$  and $x\ast \psi_{\lambda^\prime}$.

\subsection{Subset of auto-WPH moments}
\label{app:wph_subset}

\cite{Allys2020} identified a relevant set of auto-WPH moments to build models of simulated data of the large-scale structure of the Universe. In the present work, we define WPH statistics that are directly inspired by this work, although they are made slightly more optimal by taking out coefficients that were redundant. We will consider in the following five categories of moments defined as follows:
\begin{itemize}
    \item the $S^{(1,1)}$ moments, of the form $C_{\lambda,1,\lambda,1}(\ve{\tau}) = \Cov{X \ast \psi_{\lambda}(\ve{r})}{X \ast \psi_{\lambda}(\ve{r} + \ve{\tau})}$, at every $\ve{\tau}_{n, \alpha}$. They measure weighted averages of the power spectrum over the bandpass of $\psi_\lambda$ (see Eq.~\eqref{eq:wph_ps}).
    \item the $S^{(0,0)}$ moments, of the form $C_{\lambda,0,\lambda,0}(\ve{\tau})= \Cov{|X \ast \psi_{\lambda}(\ve{r})|}{|X \ast \psi_{\lambda}(\ve{r} + \ve{\tau})|}$, at every $\ve{\tau}_{n, \alpha}$. They capture information related to the sparsity of the data in the bandpass of $\psi_\lambda$.
    \item the $S^{(0,1)}$ moments, of the form $C_{\lambda,0,\lambda,1}(\ve{\tau}) = \Cov{|X \ast \psi_{\lambda}(\ve{r})|}{X \ast \psi_{\lambda}(\ve{r} + \ve{\tau})}$, at $\ve{\tau} = 0$ only. They capture information related to the couplings between the scales included in the same bandpass.
    \item the $C^{(0,1)}$ moments, of the form $C_{\lambda,0,\lambda^\prime,1}(\ve{\tau}) = \Cov{|X \ast \psi_{\lambda}(\ve{r})|}{X \ast \psi_{\lambda^\prime}(\ve{r} + \ve{\tau})}$, considering ${0 \leq j < j^\prime \leq J - 1}$, at every $\ve{\tau}_{n, \alpha}$ when $\theta = \theta^\prime$ and at $\ve{\tau} = 0$ only when $\theta \neq \theta^\prime$. They capture information related to the correlation between local levels of oscillation for the scales in the bandpasses associated with $\psi_{\lambda}$ and $\psi_{\lambda^\prime}$.
    \item the $C^{\rm phase}$ moments, of the form $C_{\lambda,1,\lambda^\prime, p^\prime}(\ve{\tau}) = \Cov{X \ast \psi_{\lambda}(\ve{r})}{\left[X \ast \psi_{\lambda^\prime}(\ve{r} + \ve{\tau})\right]^{p^\prime}}$ with ${p^\prime = 2^{j^\prime - j} > 1}$, considering $0 \leq j < j^\prime \leq J - 1$ and $\theta = \theta^\prime$, at every $\ve{\tau}_{n, \alpha}$. They capture information related to the statistical phase alignment of oscillations between the scales in the bandpasses associated with $\psi_{\lambda}$ and $\psi_{\lambda^\prime}$.
\end{itemize}

\paragraph{Symmetries in the subset of moments}

We identify in the following symmetries with respect to the angular variables that will allow us to define optimal ranges for the angular variables, avoiding redundancy in the statistical content of the WPH statistics. We denote, e.g., by $S^{(1, 1)}(j, \theta, n, \alpha)$ the moment $C_{\lambda, 1, \lambda, 1}(\ve{\tau}_{n, \alpha})$ with $\lambda = (j, \theta)$, and use similar notations for the other categories of moments.

For a real-valued statistically homogeneous random field $X$, and for a choice of wavelets satisfying the symmetry $\psi_{j, \theta + \pi} = \conjb{\psi_{j, \theta}}$ (valid when $\hat{\psi}(\ve{k}) \in \mathbb{R}$ for all $\ve{k}$, which is the case of bump-steerable wavelets), we identify the following list of symmetries:
\begin{align}
S^{(1, 1)}(j, \theta + \pi, n, \alpha) &= S^{(1, 1)}(j, \theta, n, \alpha), \\
S^{(1, 1)}(j, \theta, n, \alpha + \pi) &= \conjb{S^{(1, 1)}(j, \theta, n, \alpha)}, \label{eq:s11_alpha_sym}\\
S^{(0, 0)}(j, \theta + \pi, n, \alpha) &= S^{(0, 0)}(j, \theta, n, \alpha), \\
S^{(0, 0)}(j, \theta, n, \alpha + \pi) &= S^{(0, 0)}(j, \theta, n, \alpha), \label{eq:s00_alpha_sym}\\
S^{(0, 1)}(j, \theta + \pi) &= \conjb{S^{(0, 1)}(j, \theta)}, \\
C^{(0, 1)}(j_1, \theta_1 + \pi, j_2, \theta_2, n, \alpha) &= C^{(0, 1)}(j_1, \theta_1, j_2, \theta_2, n, \alpha), \\
C^{(0, 1)}(j_1, \theta_1, j_2, \theta_2 + \pi, n, \alpha) &= \conjb{C^{(0, 1)}(j_1, \theta_1, j_2, \theta_2, n, \alpha)}, \\
C^{\rm phase}(j_1, \theta + \pi, j_2, \theta + \pi, n, \alpha) &= \conjb{C^{\rm phase}(j_1, \theta, j_2, \theta, n, \alpha)}.
\end{align}
Proofs of these relations essentially stem from the facts that $\Cov{X}{Y} = \conjb{\Cov{Y}{X}}$ and that ${\left[ X \ast \psi_{j, \theta + \pi} \right]^p = \conjb{\left[ X \ast \psi_{j, \theta} \right]^p}}$ for any $p\in\mathbb{R}$. The latter relation is a consequence of the commutativity of the complex conjugation with the phase harmonic and convolution operations.

Coefficients that are either equal or related by a complex conjugation operation are said to be redundant. The previous relations show that, in order to avoid redundancy, it is sufficient to consider wavelets with $\theta \in [0, \pi)$. Moreover, to avoid further redundancy, for the $S^{(1,1)}$ and $S^{(0,0)}$ moments, it is sufficient to consider $\ve{\tau}_{n, \alpha}$ vectors with $\alpha\in[0, \pi)$ only. We show in Table~\ref{table:wph_nb_coeffs} the resulting number of auto coefficients per class of moments for the parameters used in this work.

\begin{table}
    \def\arraystretch{1.5}
    \centering
    \begin{tabular}{c|ccccccccc|c|c}
        \hline
        \hline
        &$S^{(1,1)}$ & $S^{(0,0)}$ & $S^{(0,1)}$ & $S^{(1,0)}$ & $C^{(0,1)}$ & $C^{(1,0)}$ & $C^{\rm phase}$ & $C^{\rm phase, inv}$ & $L$ & Total & Ratio (\%) \\
        \hline
           auto & 544 & 544 & 32 & N/A & 4032 & N/A & 1776 & N/A & 12 & 6940 & $\sim 2.6$ \\ 
           cross & 32 & 32 & 32 & 32 & 448 & 448 & 112 & 112 & 16 & 1264 & $\sim 0.24$ \\ 
        \hline
    \end{tabular}
    \caption{Number of statistical coefficients per class of moments for the parameters used in this work. The last column gives the ratio of the total number of coefficients to the number of pixels in one (or two, for the cross case) $512\times512$ image(s).}
    \label{table:wph_nb_coeffs}
\end{table}

\subsection{Subset of cross-WPH moments}
\label{app:cross_wph_subset}

Just like before, we focus on a specific subset of cross-WPH moments. We make sure that this subset is non-redundant and \textit{pseudo}-symmetric under the exchange of $X$ and $Y$ (modulo complex conjugation). These moments are the following:
\begin{itemize}
    \item the $S^{(1,1)}_\times$ moments, of the form $C_{\lambda,1,\lambda,1}^\times(\ve{\tau}) = \Covb{X \ast \psi_{\lambda}(\ve{r})}{Y \ast \psi_{\lambda}(\ve{r} + \ve{\tau})}$. Note that they measure weighted averages of the cross spectrum over the bandpass of $\psi_\lambda$.
    \item the $S^{(0,0)}_\times$ moments, of the form $C_{\lambda,0,\lambda,0}^\times(\ve{\tau})= \Covb{|X \ast \psi_{\lambda}(\ve{r})|}{|Y \ast \psi_{\lambda}(\ve{r} + \ve{\tau})|}$.
    \item the $S^{(0,1)}_\times$ moments, of the form $C_{\lambda,0,\lambda,1}^\times(\ve{\tau}) = \Covb{|X \ast \psi_{\lambda}(\ve{r})|}{Y \ast \psi_{\lambda}(\ve{r} + \ve{\tau})}$.
    \item the $S^{(1,0)}_\times$ moments, of the form $C_{\lambda,1,\lambda,0}^\times(\ve{\tau}) = \Covb{X \ast \psi_{\lambda}(\ve{r})}{|Y \ast \psi_{\lambda}(\ve{r} + \ve{\tau})|}$.
    \item the $C^{(0,1)}_\times$ moments, of the form $C_{\lambda,0,\lambda^\prime,1}^\times(\ve{\tau}) = \Covb{|X \ast \psi_{\lambda}(\ve{r})|}{Y \ast \psi_{\lambda^\prime}(\ve{r} + \ve{\tau})}$, considering ${0 \leq j < j^\prime \leq J - 1}$.
    \item the $C^{(1,0)}_\times$ moments, of the form $C_{\lambda,0,\lambda^\prime,1}^\times(\ve{\tau}) = \Covb{X \ast \psi_{\lambda}(\ve{r})}{|Y \ast \psi_{\lambda^\prime}(\ve{r} + \ve{\tau})|}$, considering ${0 \leq j^\prime < j \leq J - 1}$.
    \item the $C^{\rm phase}_\times$ moments, of the form $C_{\lambda,1,\lambda^\prime, p^\prime}^\times(\ve{\tau}) = \Covb{X \ast \psi_{\lambda}(\ve{r})}{\left[Y \ast \psi_{\lambda^\prime}(\ve{r} + \ve{\tau})\right]^{p^\prime}}$ with ${p^\prime = 2^{j^\prime - j} > 1}$, considering $0 \leq j < j^\prime \leq J - 1$ and $\theta = \theta^\prime$.
    \item the $C^{\rm inv, phase}_\times$ moments, of the form $C_{\lambda,p,\lambda^\prime, 1}^\times(\ve{\tau}) = \Covb{\left[X \ast \psi_{\lambda}(\ve{r})\right]^{p}}{Y \ast \psi_{\lambda^\prime}(\ve{r} + \ve{\tau})}$ with ${p = 2^{j - j^\prime} > 1}$, considering $0 \leq j^\prime < j \leq J - 1$ and $\theta = \theta^\prime$.
\end{itemize}
Moreover, for this work, inspired by \cite{Brochard2022}, we only consider cross-WPH moments with $\ve{\tau}=\ve{0}$. We show in Table~\ref{table:wph_nb_coeffs} the resulting number of cross coefficients per class of moments for the parameters used in this work.

\subsection{Normalized estimates}
\label{app:wph_normalization}

In practice, our statistical coefficients are all normalized similarly to \cite{Zhang2021} and \cite{Allys2020}. This normalization has been shown to speed up the minimization involved during the sampling process of microcanonical gradient descent models~\citep{Zhang2021}.

The normalization of the auto-WPH and scaling coefficients depends on the target map $x_0$ involved in this minimization. Denoting by $\tilde{C}_{\lambda,p,\lambda^\prime,p^\prime}(\ve{\tau})$ and $\tilde{L}_{j,p, p^\prime}$ the normalized estimates of $C_{\lambda,p,\lambda^\prime,p^\prime}(\ve{\tau})$ and $L_{j,p, p^\prime}$, respectively, we choose:
\begin{align}
    \tilde{C}_{\lambda,p,\lambda^\prime,p^\prime}(\ve{\tau}) &= \frac{\big \langle x^{(\lambda, p)}\left(\ve{r}\right)\conjb{x^{(\lambda^\prime, p^\prime)}}\left(\ve{r}+\ve{\tau}\right)\big \rangle}{\sqrt{\big \langle\lvert{x_0^{(\lambda, p)}}\rvert^2\big \rangle\big \langle\lvert{x_0^{(\lambda^\prime, p^\prime)}}\rvert^2\big \rangle}},\\
    \tilde{L}_{j,p, p^\prime} &= \frac{ \langle x^{(j,p)}x^{(j,p^\prime)} \rangle}{\sqrt{\langle \lvert x_0^{(j,p)}\rvert^2 \rangle\langle \lvert x_0^{(j,p^\prime)}\rvert^2 \rangle}},
\end{align}
where the brackets stand for a spatial mean on $\ve{r}$, and ${x^{(\ve{\xi}, p)} = [x \ast \psi_{\ve{\xi}}]^{p} - \big \langle[x_0 \ast \psi_{\ve{\xi}})]^{p}\big \rangle}$. Note that this definition of $x^{(\ve{\xi}, p)}$ depends on $x_0$.

For cross coefficients, we define similarly:
\begin{align}
    \tilde{C}_{\lambda,p,\lambda^\prime,p^\prime}^\times(\ve{\tau}) &= \frac{\big \langle x^{(\lambda, p)}\left(\ve{r}\right)\conjb{y^{(\lambda^\prime, p^\prime)}}\left(\ve{r}+\ve{\tau}\right)\big \rangle}{\sqrt{\big \langle\lvert{x_0^{(\lambda, p)}}\rvert^2\big \rangle\big \langle\lvert{y_0^{(\lambda^\prime, p^\prime)}}\rvert^2\big \rangle}},\\
    \tilde{L}_{j,p, p^\prime}^\times &= \frac{ \langle x^{(j,p)}y^{(j,p^\prime)} \rangle}{\sqrt{\langle \lvert x_0^{(j,p)}\rvert^2 \rangle\langle \lvert y_0^{(j,p^\prime)}\rvert^2 \rangle}}.
\end{align}

\section{Gaussian model}
\label{app:app_gaussian}

\begin{figure*}
    \centering
    \includegraphics[width=0.85\hsize]{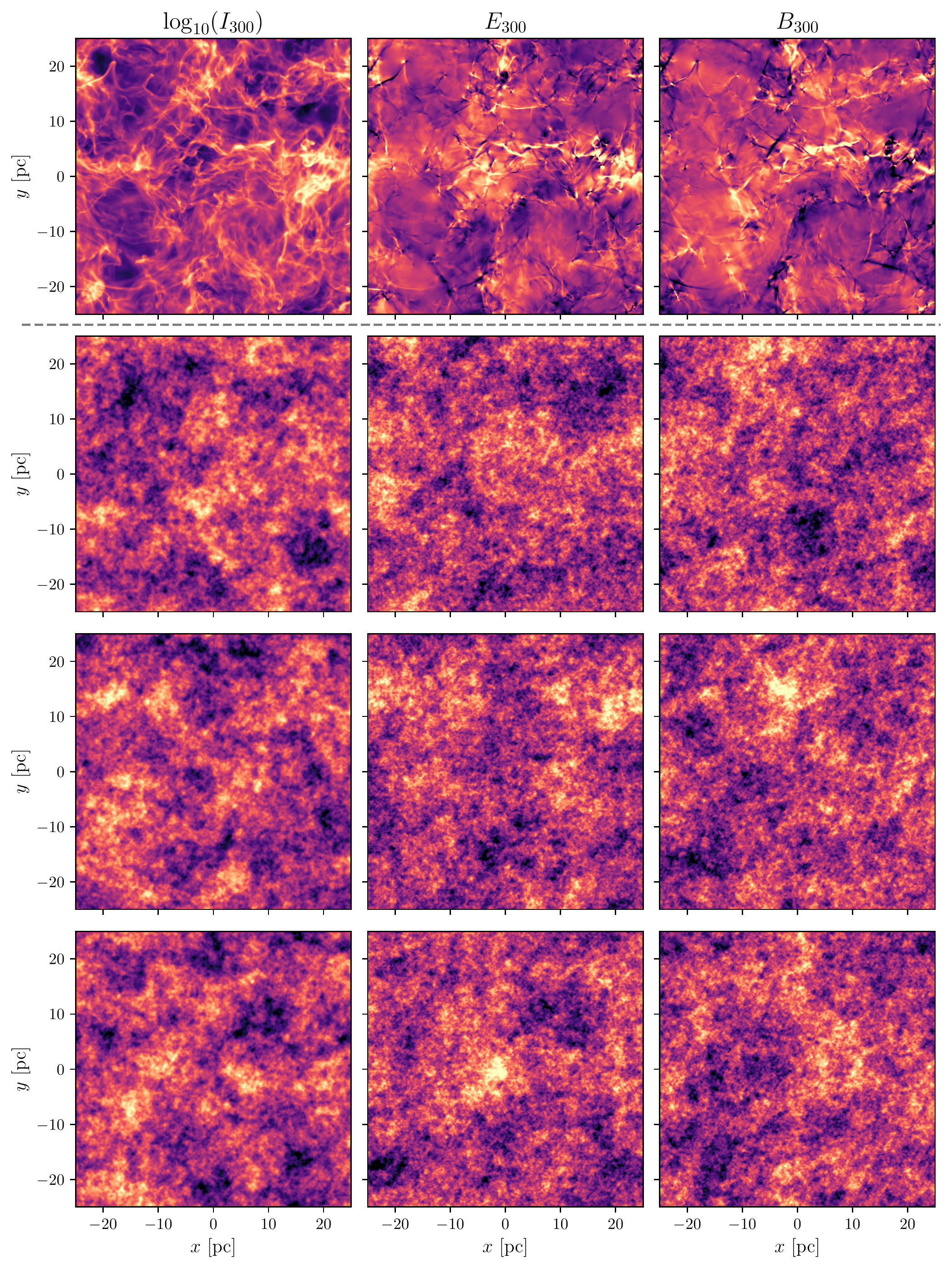}
    \caption{Same as Fig.~\ref{fig:synth_IEB_monofreq_maps} but for the Gaussian model.}
    \label{fig:synth_IEB_monofreq_gaussian_maps}
\end{figure*}

The Gaussian model of the data $\ve{x} = (\log(I_{300}), E_{300}, B_{300})$, which is used as a baseline in Sect.~\ref{sec:monofreq_model}, is defined as follows. It is a microcanonical model conditioned by:
\begin{equation}
    \phi^\mathrm{G}(\ve{x}) = \phi_{\rm auto}^\mathrm{G}(\log(I_{300})) \oplus \phi_{\rm auto}^\mathrm{G}(E_{300}) \oplus \phi_{\rm auto}^\mathrm{G}(B_{300}) \oplus \phi_{\rm cross}^\mathrm{G}(\log(I_{300}), E_{300}) \oplus \phi_{\rm cross}^\mathrm{G}(\log(I_{300}), B_{300})),
\end{equation}
where $\phi_{\rm auto}^\mathrm{G}(\cdot)$ and $\phi_{\rm cross}^\mathrm{G}(\cdot, \cdot)$ only include normalized estimates of the $S^{(1, 1)}$ and $S^{(1, 1)}_\times$ moments as introduced in Sect.~\ref{app:wph_subset} and \ref{app:cross_wph_subset}, respectively. Such microcanonical models are rigorously Gaussian~\citep{Bruna2019}. We show in Fig.~\ref{fig:synth_IEB_monofreq_gaussian_maps} three different samples of this model (from second to last row) next to the original data $\ve{x}$ (first row).

\end{appendix}